\documentclass[aps,prd,10pt,nofootinbib,twocolumn,eqsecnum,showpacs,showkeys,superscriptaddress,preprintnumbers,altaffilletter]{revtex4-1}

\pdfoutput=1
\usepackage{graphicx}
\usepackage{dcolumn}
\usepackage{amssymb,amsmath}
\usepackage{amsfonts}
\usepackage{amsbsy}
\usepackage{color}
\usepackage{rotating}
\usepackage[english]{babel}
\usepackage{multirow}
\usepackage{hyperref}
\hypersetup{
    colorlinks=true,
    linkcolor=blue,
    filecolor=magenta,
    urlcolor=cyan,
    citecolor=cyan
}

\newcommand{\lp}{\ell_{\rm p}}
\newcommand{\mpl}{m_{\rm p}}

\newcommand{\ag}{A_{\rm G}}
\newcommand{\kb}{k_{\rm B}}
\newcommand{\fg}{f_{\rm G}}
\newcommand{\ra}{\tilde{r}_{\mathrm{A}}}
\newcommand{\sg}{S_{\rm G}}
\usepackage{lmodern}
\usepackage{sansmath}
\numberwithin{equation}{section}
\newcommand{\alphaq}{\alpha_{\rm Q}}
\newcommand{\alphal}{\alpha_{\rm L}}

\begin{document}

\title{Cosmological constraints on the Generalized Uncertainty Principle from modified Friedmann equations}

\author{Serena Giardino}
\email{serena.giardino@usz.edu.pl}
\affiliation{Institute of Physics, University of Szczecin, Wielkopolska 15, 70-451 Szczecin, Poland}
\author{Vincenzo Salzano}
\email{vincenzo.salzano@usz.edu.pl}
\affiliation{Institute of Physics, University of Szczecin, Wielkopolska 15, 70-451 Szczecin, Poland}

\date{\today}

\begin{abstract}
The Generalized Uncertainty Principle (GUP) has emerged in numerous attempts to a theory of quantum gravity and predicts the existence of a minimum length in Nature. In this work, we consider two cosmological models arising from Friedmann equations modified by the GUP (in its linear and quadratic formulations) and compare them with observational data. Our aim is to derive constraints on the GUP parameter and discuss the viability and physical implications of such models.
We find for the parameter in the quadratic formulation the constraint $\alpha^{2}_{Q}<10^{59}$ (tighter than most of those obtained in an astrophysical context) while the linear formulation does not appear compatible with present cosmological data. Our analysis highlights the powerful role of high-precision cosmological probes in the realm of quantum gravity phenomenology.
\end{abstract}


\maketitle

\section{Introduction}
The presence of a Generalized Uncertainty Principle (GUP) appears to be an ubiquitous feature in different tentative approaches to quantum gravity and thought experiments. This modification of the Heisenberg Uncertainty Principle (HUP) is required whenever the gravitational interaction is taken into account and results in the introduction of a minimum length scale in Nature, of the order of Planck length. The existence of such a minimum length has powerful implications, since it entails the impossibility to probe lengths shorter than Planck length \textit{in principle}, regardless of the capabilities of any experimental apparatus, which seems to deeply affect the very concept of spacetime.

The idea of a minimum length in Nature has a long history \cite{hossenfelder}, but only in the '60s, the pioneering work of Mead \cite{mead} pointed out the crucial role of the gravitational interaction in the possibility of probing very short distances. However, it is mostly after Hawking's revolutionary insights into the thermodynamical properties of black holes (BHs) \cite{hawking} that the ``trans-Planckian problem'' has emerged as a relevant issue for gravity (and, after that, also for inflationary cosmology \cite{brandenberger}).
In the following years, results in String Theory brought more attention to the problem of resolving infinitely small lengths with extended objects \cite{amati, veneziano, adler}. Since then, the GUP has played an important role as a heuristic tool to understand quantum gravity effects, starting from \textit{Gedankenexperimente} and disparate theoretical frameworks, even if the GUP itself seems to nonetheless have model-independent features \cite{maggiore}.

One of the contexts in which the GUP has been studied most extensively is that of BHs: as Bronstein had already understood back in the 1930s \cite{bronstein}, the problem of finding a theory of quantum gravity is tightly interwoven with the existence of BHs. As gravity does not allow the concentration of an infinite amount of energy into a confined region, because the process will end in gravitational collapse and the formation of a BH, there is an inescapable limit to the precision to which the gravitational field can be measured \cite{gorelik}.

This insight is remarkably similar to the \textit{Gedankenexperiment} involving the formation of micro black holes carried out in \cite{scardigli}, which yields an intuitive explanation of the intrinsic limit to any measurement carried out around the Planck scale.

The most common expression of the GUP is given by
\begin{equation}
    \Delta x\Delta p \geq \dfrac{\hbar}{2}\left(1+\beta\dfrac{\lp^2}{\hbar^2}\Delta p ^2\right),
    \label{gupbasis}
\end{equation}
where $\lp=1.61623\cdot 10^{-35}\,\rm{m}$ is the Planck length and $\beta$ is the dimensionless GUP parameter, generically assumed to be $\mathcal{O}(1)$, because the minimum length arising from the GUP is $\Delta x_{\rm min}\approx \sqrt{\beta}\lp$ \cite{maggiore}. Here, the correction to HUP is proportional to a quadratic term in the momentum uncertainty, but several other formulations of the GUP also contain a linear term \cite{discrete}.

GUP-induced quantum effects lead to corrections both of the Hawking temperature and the Bekenstein entropy. Because of such modifications and their influence on the flux of Hawking radiation \cite{anamariu,Scardigli:2016pjs}, the GUP critically affects BH evaporation \cite{adlerchen, scardcasa}.
Especially in the final stages of this process, the role of the GUP becomes important as it could cause the black hole to leave a remnant of Planck size, thus providing hints towards the resolution of the black hole information paradox.

An additional intriguing implication is that the correction to the entropy, translated to a modified entropy-area law \cite{Bekenstein:1972tm,Bekenstein:1974ax,Hawking:1974rv,vagenas,anacleto}, can also be studied in a cosmological context because of the geometrical (and thus universal) nature of such a law, which does not only apply to black hole horizons.
Crucially, such results are general because any causal horizon is inevitably associated with entropy, since by definition it hides information from observers, as clarified in the seminal paper by Jacobson \cite{jacobson}.
Showing that the first law of thermodynamics can be recast as the Einstein equations, the author provided evidence of the close relationship between thermodynamics and gravity, resulting in illuminating insights into the thermodynamical properties of space-time as a whole.

Building on this result, the authors of \cite{caikim} showed that the Friedmann equations can be recovered by applying the first law of thermodynamics, $dE=TdS$, to the apparent horizon of Friedmann-Lema\^itre-Robertson-Walker (FLRW) spacetime, which is also endowed with a temperature $T$ and an entropy $S$ that read
\begin{equation}
  T=\frac{1}{2\pi \ra}\; \quad\quad  S=\frac{A}{4G},
  \label{tes}
\end{equation}
where $A$ is the area and $\ra$ the radius of the apparent horizon.

This procedure for obtaining the Friedmann equations seems to have quite broad validity: it was shown that it also holds for alternative theories of gravity \cite{caikim} and even if the entropy-area law is generically modified \cite{caicao}.

More recently, the authors of \cite{renli} have shown that the Friedmann equations are still recovered (albeit in a modified form) even if the entropy-area law is affected by the GUP, which means that GUP-induced quantum effects at high energies can indeed influence the dynamics of the FLRW universe at early times, however slightly.

In addition to the plethora of theoretical investigations carried out on the GUP, a research direction rooted in quantum gravity phenomenology is attempting to quantify the magnitude of GUP-induced quantum corrections by constraining the GUP parameter. The relevance of these studies lies in the fact that they open a much needed low-energy window on quantum gravity, far from the presently inaccessible Planck scale, employing precision experiments in many areas of physics.
However, the instances in which theoretical studies are complemented by comparison with experimental data, especially in a cosmological setting, are, to the best of our knowledge, scarce (exceptions are e.g.\ \cite{kouwn, leandros}).

The aim of the present work is precisely to bridge two cosmological models in which GUP-induced thermodynamical corrections are taken into account \cite{renli, majumder} with the wealth of precise cosmological data available today. Our aim is two-fold: on the one hand, we will check whether these models are compatible with data and thus study their cosmological viability; on the other hand, we will derive cosmologically-motivated constraints on the GUP parameter. Generally speaking, as clarified in Section \ref{bounds}, experimental bounds coming from the study of gravitational effects are weak, while bounds coming from quantum experiments are much more stringent. We will show that cosmological data are capable of providing fairly strong constraints, comparable in magnitude with the less stringent estimates of quantum experiments.

The paper is organized as follows: in Sections \ref{background} and \ref{bounds}, we will review the salient steps for the derivation of the modified Friedmann equations and the experimental bounds on the GUP parameter; in Section \ref{mcmc}, we will describe the statistical methods and cosmological data employed in our analysis; in section \ref{results} we will present and discuss our results.

\section{Theoretical background}
\label{background}

The work of \cite{caikim} characterises the apparent horizon of FLRW universe in analogy with the event horizon of a BH. The apparent horizon is a marginally trapped surface with vanishing expansion that always exists in FLRW universe (differently from the event and particle horizons) which makes it the best suited cosmological horizon for thermodynamical considerations, also in view of its trapping character \cite{faraoni}.
To set our notation, we write the Friedmann equations as
\begin{eqnarray}
      H^2&=&\frac{8\pi G}{3}\rho-\frac{kc^2}{a^2} \\
      \dot{H}&=&-4\pi G\left(\rho+\frac{p}{c^2}\right) + \frac{kc^2}{a^2}. \nonumber
\end{eqnarray}
The continuity equation reads
\begin{equation}
    \dot{\rho}+3H\left(\rho+\frac{p}{c^2}\right)=0
    \label{contin}
\end{equation}
and will not be affected by the GUP corrections.

Since we are interested in recovering the GUP-modified entropy-area law, we need to define the radius and the area of the apparent horizon. The radius is
\begin{equation}
    \ra=\frac{c}{\sqrt{H^2+\frac{kc^2}{a^2}}}
\label{ra}
\end{equation}
and yields the area
\begin{equation}
    A=4\pi\ra^2=\frac{4\pi c^2}{H^2+\dfrac{kc^2}{a^2}}.
    \label{a}
\end{equation}

The expressions for the entropy and the temperature associated to a BH horizon read
\begin{equation}
S=\frac{\kb c^3 A}{4G\hbar}\; \quad\quad T=\frac{\hbar c^3}{8\pi G \kb M},
\label{set}
\end{equation}
where $A$ is the area of the event horizon, $M$ the mass of the black hole and $k_{\rm B}$ the Boltzmann constant. In the well-studied case of a Schwarzschild BH, the Schwarzschild radius $r_{\rm S}=2GM/c^2$ is used to obtain
\begin{equation}
    T=\frac{\hbar c}{4\pi \kb r_{\rm S}}.
\end{equation}
In \cite{gibbhawk}, this result has been generalized to the case of a de Sitter universe, and in \cite{caikim} it is only \textit{assumed} as a working hypothesis that the expressions above also work for the apparent horizon. However, since the authors do successfully recover the Friedmann equations, their assumption is justified \textit{\textit{a posteriori}}.
Thus, for our purposes, we can simply assume \cite{caihawking}
\begin{equation}
    T=\frac{\hbar c}{4\pi \kb\ra}.
\end{equation}


\subsection{Quadratic GUP}
We will now briefly review the treatment of \cite{renli}, in which it is shown in an arbitrary number of dimensions that a modified form of the Friedmann equations can be recovered when the GUP corrects the standard entropy-area law \eqref{tes}.
In the following, we will also restore all physical units which are generally omitted in theoretical studies, since our goal is to find a numerical estimate for the GUP parameter.
The authors of \cite{renli} start from the following expression for the GUP, which includes a quadratic correction in momentum uncertainty (note that the dimensionless GUP parameter $\beta$ in \eqref{gupbasis} is here $\alphaq^2$, where $Q$ stands for ``quadratic'')
\begin{equation}
    \Delta x \Delta p \geq \frac{\hbar}{2}\left( 1 +\frac{\alphaq^{2} \lp^2}{\hbar^2}\Delta p^2 \right).
    \label{quadgup}
\end{equation}
We can straightforwardly solve for the momentum uncertainty, obtaining
\begin{equation}
    \Delta p\geq\frac{\hbar\, \Delta x}{\alphaq^2\lp^2}-\sqrt{\frac{\hbar^2\,\Delta x^2}{\alphaq^4\lp^4}-\frac{\hbar^2}{\alphaq^2\lp^2}},
\end{equation}
which can be recast as
\begin{equation}
    \Delta p\geq\frac{\hbar}{2 \Delta x}\left[\frac{2\Delta x^2}{\alphaq^2\lp^2}-\frac{2 \Delta x^2}{\alphaq^2\lp^2}\sqrt{1-\frac{\alphaq^2\lp^2}{\Delta x^2}}\right].
\end{equation}
The expression inside square brackets is the function characterizing the departure of the GUP from the HUP, which we define as
\begin{equation}
    \fg(\Delta x^2)=\frac{2\Delta x^2}{\alphaq^2\lp^2}-\frac{2\Delta x^2}{\alphaq^2\lp^2}\sqrt{1-\frac{\alphaq^2\lp^2}{\Delta x^2}}.
\label{fg}
\end{equation}


In order to study the GUP effects on the thermodynamics of FLRW universe, the authors of \cite{renli} (relying on \cite{park}) consider the following picture: if the apparent horizon has absorbed or radiated a particle with energy $dE$, this energy can be identified with the uncertainty in momentum, $dE\simeq c\Delta p$. Therefore, the HUP $\Delta p\geq\hbar/(2\Delta x)$ yields the corresponding increase or decrease in the area of the apparent horizon, due to \eqref{set}:
\begin{equation}
  dA=\frac{4G\hbar}{\kb c^3 T}dE\simeq\frac{2G\hbar^2}{\kb c^2 T\Delta x}.
  \label{da}
\end{equation}
However, in the case where the GUP is taken into account, this relationship becomes
\begin{equation}
    d\ag\simeq\frac{2G\hbar^2}{\kb c^2 T}\frac{\fg(\Delta x^2)}{\Delta x},
\end{equation}
with
\begin{equation}
    d\ag=\fg(\Delta x^2)dA.
    \label{dag}
\end{equation}

The uncertainty in position of the absorbed or radiated particle is reasonably considered of the order of its Compton length $\lambda$, which is approximately the inverse of the Hawking temperature in natural units. In physical units, as noted in \cite{vagenas}, it is customary to consider $\Delta x\simeq\lambda$.
For a Schwarzschild BH, the particle has a wavelength of the order of the inverse Hawking temperature (for an asymptotic observer) or, more generally, of the inverse of the surface gravity $\kappa^{-1}=2r_{\rm S}$ (since $T=\kappa/2\pi$ in natural units). As previously noted, it seems sensible to extend the argument from the context of BHs to that of the apparent cosmological horizon, thus assuming $\Delta x\simeq2\ra=\sqrt{A/\pi}$.

We can express the departure function \eqref{fg} in terms of the area of the apparent horizon and subsequently in terms of the entropy. If we expand $\fg(A)$ around $\alphaq\, l_p =0$, we obtain
\begin{equation}
    \fg(A)=1+\frac{\pi\alphaq^2\lp^2}{4}\frac{1}{A}+\frac{\pi^2\alphaq^4\lp^4}{8}\frac{1}{A^2}+\mathcal{O}\left(\alphaq^6\right)
\label{fga}
\end{equation}
up to second order, which is sufficient for the purposes of this work. Substituting \eqref{fga} in \eqref{dag} and integrating, we find the expression
\begin{equation}
    \ag=A+\frac{\pi\alphaq^2\lp^2}{4} \ln{A}-\frac{\pi^2\alphaq^4\lp^4}{8}\frac{1}{A},
    \label{ag}
\end{equation}
where in $\ln{A}$ we have included the integration constant $A_0$.

For the entropy, due to \eqref{dag} and \eqref{set}, we obtain $d\sg=\fg(A)\,dS$.
After integration, we find the modified entropy-area relation to be
\begin{equation}
       \sg=\frac{\kb c^3}{4G\hbar}\left[A+\frac{\pi\alpha^2\lp^2}{4} \ln{A}-\frac{\pi^2\alpha^4\lp^4}{8}\frac{1}{A}\right].
\label{entropyarea}
\end{equation}

The apparent horizon approach to find the Friedmann equations devised in \cite{caikim} consists in the application of the first law of thermodynamics to the apparent horizon of a FLRW universe, with the additional assumption \eqref{set}.
This procedure involves the definition of a work density and an energy supply vector, respectively regarded as the work done by a change of the apparent horizon and the total energy flow through it, which is associated to the entropy. Such definitions yield a specific form of the first law of thermodynamics for cosmological horizons.
The first time derivative of \eqref{ra} is
\begin{equation}
    \dot{\tilde{r}}_{\mathrm{A}}=-\frac{1}{c^2}\ra^3 H\left(\dot{H}-\frac{kc^2}{a^2}\right),
\end{equation}
which can be rewritten as
\begin{equation}
    \frac{d\ra}{\ra^3}=-\frac{1}{c^2}H\left(\dot{H}-\frac{kc^2}{a^2}\right)dt.
    \label{dra}
\end{equation}

Taking into account \eqref{fg} and \eqref{entropyarea}, it is straightforward to see that
\begin{equation}
    \sg'(A)=\frac{\kb c^3}{4G\hbar}\fg(A),
    \label{sgprime}
\end{equation}
where $\sg'$ is a first derivative with respect to $A$.

Following \cite{caicao} and \cite{awadali}, we can use
\begin{equation}
    \fg(A)\frac{d\ra}{\ra^3}=\frac{4\pi G}{c^2}\left(\rho+\frac{p}{c^2}\right)H dt.
    \label{fdr}
\end{equation}
as the starting point for finding the Friedmann equations.

On the one hand, using \eqref{dra}, we can find the dynamical Friedmann equation
\begin{equation}
    \fg(A)\left(\dot{H}-\frac{kc^2}{a^2}\right)=-4\pi G \left(\rho+\frac{p}{c^2}\right).
    \label{drr}
\end{equation}
On the other hand, taking into account the continuity equation \eqref{contin}, \eqref{fdr} can be recast as
\begin{equation}
     \frac{8\pi G}{3}d\rho=-4\pi c^2 \fg(A)\frac{dA}{A^2}.
     \label{pre1}
\end{equation}
Integrating this equation yields the Friedmann constraint, so that, in summary, the two GUP-modified Friedmann equations (up to second order in $\alphaq^2$) are given by
\begin{equation}
\frac{8\pi G\rho}{3}=4\pi c^2\left[\frac{1}{A}+\frac{\pi\alphaq^2\lp^2}{8}\frac{1}{A^2}+\mathcal{O}\left(\alphaq^4\right)\right]
\label{fried1}
\end{equation}
\begin{eqnarray}
\label{fried2}
&&-4\pi G\left(\rho+\frac{p}{c^2}\right) = \left(\dot{H}-\frac{kc^2}{a^2}\right)\cdot \\
&&\cdot\left[1+\frac{\pi\alphaq^2\lp^2}{4}\frac{1}{A}+\frac{\pi^2\alphaq^4\lp^4}{8}\frac{1}{A^2}+\mathcal{O}\left(\alphaq^4\right)\right]. \nonumber
\end{eqnarray}

\subsection{Implementing the Friedmann equations}
\label{implementing}
In order to compare the model arising from these modified Friedmann equations with cosmological data, we need an expression
for $H(z)$ in terms of the cosmological parameters. Our cosmological background will be a standard $\Lambda$CMD model \cite{Bull:2015stt}, fully characterized by the Hubble constant $H_0=100 \cdot h$, the dimensionless density parameters for matter $\Omega_{m}$, for radiation $\Omega_{r}$ and for a cosmological constant as dark energy component $\Omega_{\Lambda}$, where $\Omega_i=8\pi G\rho_i/3H_0^2$ and the curvature $k$ is expressed as $\Omega_{ k}=-kc^2/H_0^2$.

As a first step, we solve equation \eqref{fried1} for $A$, choosing the positive and real solution
\begin{equation}
    A=\frac{3 c^2+\sqrt{3} \sqrt{3 c^4+ \pi c^2 \alphaq ^2\lp^2 G \rho}}{4 G \rho }.
\end{equation}
Due to \eqref{a}, we can invert to find
\begin{equation}
   H^2(z)=\frac{16 \pi G \rho }{3 +\sqrt{3} \sqrt{3 + \pi G\rho \dfrac{\alphaq ^2 \lp^2}{c^2}}} + \Omega_{ k} (1+z)^2,
   \label{h1}
\end{equation}
where $a=1/(z+1)$ and $\rho$ is the total energy-matter density of the universe. Given the continuity equation (\ref{contin}), each component of the energy density is $\rho_{i}=\rho_{0,i}a^{-3(1+w_i)}$, where the equation of state parameter is $w_i=p_i/\rho_i$, $\Omega_i=\rho_{0,i}/\rho_{c,0}$ and  $\rho_{c,0}=3H_0^2/(8\pi G)$. Therefore, it is possible to write the total energy density $\rho$ in \eqref{h1} as $\rho=\sum_i\Omega_i\rho_{c}a^{-3(1+w_i)}$ and express the dimensionless Hubble parameter $E(z)=H(z)/H_0$ as
\begin{equation}
    E(z)=\sqrt{\frac{2 X(z)}{1+\sqrt{1+ \dfrac{H_0^2 \alpha ^2 \lp^2}{8c^2} X(z)}}+\Omega_{ k} (1+z)^2}\, ,
    \label{ez}
\end{equation}
where
\begin{equation}
X(z) = \Omega_{\Lambda}+\Omega_{ m} (1+z)^3+\Omega_{ r} (1+z)^4.
\end{equation}
Additionally, after ensuring the normalization condition $E(z=0)=1$, we can define $\Omega_{\Lambda}$ in terms of all other parameters, inverting \eqref{ez} at $z=0$, which yields
\begin{equation}
    \Omega_{\Lambda}= (1-\Omega_k-\Omega_m-\Omega_r) + \frac{H_0^2 \alpha ^2 \lp^2 \left(1-\Omega_{k}\right)^2}{32\, c^2} .
    \label{omegal}
\end{equation}

\subsection{Linear GUP}
\label{lineargup}
Making use of an alternative formulation of the GUP proposed in \cite{discrete}, the author of \cite{majumder} also performed the computation of the modified Friedmann equations with the apparent horizon formalism. More specifically, this work deals with
\begin{equation}
\Delta x\Delta p\geq \frac{\hbar}{2}\left(1+\frac{\alphal \lp}{\hbar}\Delta p +\frac{\alphal^2\lp^2}{\hbar^2}\Delta p^2\right),
\end{equation}
where, in addition to the quadratic term in the momentum, a linear term appears (the GUP parameter is named $\alphal$ here, where $L$ stands for ``linear'', to avoid any ambiguities with the quadratic case). Following the same procedure of Section~\ref{background}, the resulting modified Friedmann equations read
\begin{equation}\label{friedmaj1}
\frac{8\pi G\rho}{3}=4\pi c^2\left[\frac{1}{A}+\sqrt{\pi}\frac{\alphal\lp}{3}\frac{1}{A^{3/2}} + \mathcal{O}(\alphal^2) \right]
\end{equation}
\begin{eqnarray}
\label{friedmaj2}
&&-4\pi G\left(\rho+\frac{p}{c^2}\right) = \left(\dot{H}-\frac{kc^2}{a^2}\right)\cdot \\
&& \cdot\left[1+\sqrt{\pi}\frac{\alpha \lp}{2}\frac{1}{A^{1/2}} + \frac{\pi \alphal^2 \lp^2}{2}\frac{1}{A} + \mathcal{O}\left(\alpha^{3/2}\right)\right]
\nonumber
\end{eqnarray}
where terms containing higher orders of $1/A$ have been neglected. Given the similarity of the Friedmann equations \eqref{friedmaj1} and \eqref{friedmaj2} with \eqref{fried1} and \eqref{fried2} of \cite{renli}, it should be straightforward to compare them with cosmological data, adopting the method in Section~\eqref{implementing}.

However, this task proved substantially more challenging with the inclusion of a linear term in the GUP, due to the fractional exponent of $A$ involved in \eqref{friedmaj1}. If we assume spatial flatness, neglecting the curvature $k$, \eqref{friedmaj1} can be recast as
\begin{equation}
\frac{8\pi G\rho}{3} = H^{2} + \frac{\alphal \lp}{6c} H^{3},
\end{equation}
i.e.\, a third order equation in $H(z)$. It admits one single real solution, which reads (in terms of the dimensionless Hubble parameter $E(z)$)
\begin{equation}
    E(z)=\frac{2\,c}{\alphal\lp\,H_0}\left[\frac{F^{2}(z)-F(z)+1}{F(z)}\right],
\label{linear}
    \end{equation}
where
\begin{equation}
    F(z)=\left[\dfrac{2}{X(z)+\sqrt{-4+X^2(z)}}\right]^{1/3}
\end{equation}
and
\begin{eqnarray}
    X(z)&=&-2+ \frac{3\alphal^2 \lp^2\, H_0^2}{4c^2}\cdot \\
    &\cdot&\left(\Omega_{m}(1+z)^3+\Omega_{r}(1+z)^4+\Omega_{\Lambda}\right)\, . \nonumber
\end{eqnarray}
Expression \eqref{linear} contains a square root, which has profound implications for our goal of finding an \textit{upper} bound on the GUP parameter. Indeed, an additional condition needs to be satisfied to guarantee that $E(z)$ is real, namely
\begin{equation}
\alphal^2>\frac{16\, c^2}{3\, \lp^2 H_0^2 \left[\Omega_{m}(1+z)^3+\Omega_{r}(1+z)^4+\Omega_{\Lambda}\right]}.
\label{condi}
\end{equation}
If we restrict to positive values of the GUP parameter only, this requirement imposes a \textit{lower} bound on $\alphal$, which proves incompatible with the notion that General Relativity and standard Quantum Mechanics are to be recovered in the limit $\alphal\rightarrow0$.

An upper bound would be theoretically possible if negative values of the GUP parameter were allowed.
However, this would contrast with the idea of a minimum length $\Delta x_{\rm min}\approx \sqrt{\beta} \lp$ (for the GUP formulation \eqref{gupbasis}) which cannot be imaginary \cite{Alasfar:2017loh}. Nonetheless, a negative GUP parameter has interesting implications, see e.g.\ \cite{Jizba:2009qf,Ong:2018nzk,Buoninfante:2019fwr,Ong:2018zqn,kouwn}. For the purposes of this work, we found that allowing $\alphal$ to be negative leads to the unphysical result $H(z)<0$. This is the reason why we only consider $\alphal>0$ in order to test the model with cosmological data.

\section{Experimental bounds on GUP}
\label{bounds}

The diversity of approaches used to find experimental bounds on the GUP parameter is even richer than that of the theoretical frameworks they stem from \cite{tawfik,diab}. Still, we can clearly identify at least a couple of broad contexts in which the effects of GUP have been most consistently studied \cite{scardrev}: that of gravitational tests and that of quantum optics and atomic experiments.

In the first context, we find for example the study of a GUP-deformed Hawking temperature and Schwarzschild metric connected to Solar System tests of GR \cite{casadio} and the recent constraints on both $\alphaq$ and $\alphal$ provided by the gravitational wave event GW150914, through the analysis of modified dispersion relations for gravitons \cite{gw}. These bounds are typically quite weak, even if they seem to be tighter in approaches that violate the Equivalence Principle \cite{ghosh,Gao:2017zch}.

On the other hand, there are authors who consider the deformed commutator corresponding to the GUP,
\begin{equation}
    [\hat{X},\hat{P}]=i \hbar \left(1+\beta \dfrac{\hat{P}^2}{\mpl^2 c^2}\right),
\end{equation}
where the operators $\hat{X}$ and $\hat{P}$ are believed to be valid near Planck scale. This commutator is then used to find corrections to standard quantum mechanical effects, such as the Lamb shift and the Landau levels.
These works mainly use the GUP formulation introduced in \cite{discrete} and have provided some amongst the most stringent bounds on the GUP parameter so far \cite{qglab,Das:2008kaa,Das:2009hs}. Another development in this direction, which employs the quadratic GUP formulation, recently provided probably the tightest constraint available \cite{oscillators}.

Most of the known constraints are reported in Table \ref{tab:bounds}. The bounds on $\alphaq$ in \eqref{quadgup} and $\alphal$ in \eqref{linear} have been squared, in order to compare them with $\beta\sim\alpha_{\rm Q,L}^2$ in \eqref{gupbasis}. Nonetheless, caution may be required when contrasting bounds from completely different experiments in which the GUP influences the systems in disparate ways.

To the best of our knowledge, there are very few studies that aim to constrain the GUP parameter with cosmological observations. One of them is \cite{kouwn}, which makes use of the apparent horizon formalism to inquire whether the GUP effects could account for dark energy. However, the author mostly deals with an alternative formulation of the GUP, the General Extended Uncertainty Principle (GEUP), which predicts the existence of a maximum length as well as a minimum one. The bound this study finds on the minimum length can be regarded as a bound on $\beta$ (because $\Delta x_{\rm min}\approx\sqrt{\beta}\lp$) and is not very stringent, as reported in Table \ref{tab:bounds}. Additionally, the results of another study that uses astrophysical data \cite{gw} show that very different estimates of the GUP parameter can be found, whether the linear or the quadratic formulation is employed.

{\renewcommand{\tabcolsep}{1.5mm}
{\renewcommand{\arraystretch}{1.25}
\begin{table*}
\begin{minipage}{0.65\textwidth}
\caption{Experimental upper bounds on $\alphaq^2$ and $\alphal^2$ at $1\sigma$ confidence level.} \label{tab:bounds}
\centering
\resizebox*{\textwidth}{!}{
\begin{tabular}{lcccc}
\hline
 Experiment & $\alphaq^2$ & $\alphal^2$ & Ref. \\ 
 \hline \hline
 Harmonic oscillators & $10^6$ & $-$ & \cite{oscillators} \\
 Anomalous magnetic moment of the muon & $-$ & $10^{16}$ & \cite{Das:2011tq} \\
 Lamb shift & $10^{36}$
 & $10^{20}$ & \cite{Das:2008kaa,qglab} \\
 Scanning Tunneling Microscope & $10^{21}$ & $-$
 & \cite{Das:2008kaa} \\
 Equivalence Principle violation & $10^{21}$ & $-$ & \cite{ghosh} \\
  Weak Equivalence Principle violation & $10^{27}$ & $-$ & \cite{Gao:2017zch} \\
 Gravitational bar detectors & $10^{33}$ & $-$ & \cite{Marin:2013pga} \\
 Charmonium levels & $-$ & $10^{34}$ & \cite{qglab} \\
 Superconductivity & $-$ & $10^{34}$ & \cite{Das:2011tq} \\
 $^{87}$Rb cold-atom-recoil experiment & $10^{39}$ & $10^{28}$ & \cite{Gao:2016fmk} \\
 Landau levels & $10^{50}$ & $10^{46}$ & \cite{Das:2009hs,qglab} \\
 Gravitational waves & $10^{60}$ & $10^{40}$ & \cite{gw} \\
 Perihelion precession (Solar system data) & $10^{69}$ & $-$ & \cite{casadio} \\
 Perihelion precession (Pulsar data) & $10^{71}$ & $-$ & \cite{casadio} \\
 Light deflection & $10^{78}$ & $-$ & \cite{casadio} \\
 Cosmological observations & $10^{81}$ & $-$ & \cite{kouwn} \\
 Black hole shadow & $10^{90}$ & $-$ & \cite{neves} \\
\hline
 \textbf{Full data cosmology} & $\boldsymbol{10^{59}}$ & $-$ & \textbf{This work} \\
 \textbf{Late-time cosmology}  & $\boldsymbol{10^{81}}$ & $\boldsymbol{10^{83}}$ & \textbf{This work} \\
\hline
\hline
\end{tabular}}
\end{minipage}
\end{table*}}}

\section{Statistical Analysis}
\label{mcmc}

In order to test the viability of the models described in Section \eqref{background}, we use a combination of data coming from well-known geometrical probes. More specifically, we employ: Type Ia Supernovae (SNeIa) from the Pantheon sample; Early-Type Galaxies as Cosmic Chronometers (CC); the $H_0$ Lenses in COSMOGRAIL's Wellspring (H0LiCOW) data \cite{Wong:2019kwg}; the ``Mayflower'' sample of Gamma Ray Bursts (GRBs); Baryon Acoustic Oscillations (BAO) from several surveys; and the latest \textit{Planck} $2018$ release for Cosmic Microwave Background radiation. We consider two different scenarios: the ``full'' data set, joining both early- (CMB and BAO data from SDSS) and late-time observations (SNeIa, CC, H0LiCOW, GRBs and BAO from WiggleZ); and the ``late-time'' data set, which includes only late-time data. In general, the total $\chi^2$ is the sum of all the considered contributions 
\begin{equation}
\chi^{2}= \chi_{SN}^{2}+\chi_{G}^{2}+\chi_{H}^{2}+\chi_{H_{COW}}^{2}+ \chi_{BAO}^{2} + \chi_{CMB}^{2}\, .
\end{equation}
In order to test the predictions of the GUP-modified cosmological models given the observational data, we use our own implementation of a Monte Carlo Markov Chain (MCMC) \cite{Berg,MacKay,Neal} to minimise the total $\chi^{2}$. We test its convergence using the method developed in \cite{Dunkley:2004sv}.

As means of comparison, we also analyse a standard $\Lambda$CDM model using the same data set. This allows us to assess the reliability of the GUP-modified models with respect to the standard case, through the computation of their Bayesian Evidence, using the algorithm in \citep{Mukherjee:2005wg}. Because of the stochastic nature of the evidence, we compute it $\sim 100$ times and consider the median and the $1\sigma$ confidence level obtained from its statistical distribution, reporting them in Table \ref{tab:results}.
We briefly remark that the Bayesian Evidence $\mathcal{E}$ is defined as the probability of the data $D$ given the model $M$ with a set of parameters $\boldsymbol{p}$, i.e.\ $\mathcal{E}(M) = \int \mathrm{d}\boldsymbol{p}\ \mathcal{L}(D|\boldsymbol{p},M)\ \pi(\boldsymbol{p}|M)$, where $\pi(\boldsymbol{p}|M)$ is the prior on the set of parameters, normalised to unity, and $\mathcal{L}(D|\boldsymbol{p},M) \propto \exp -\chi^2/2$ is the likelihood function. In order to minimise the dependence on the priors \citep{Nesseris:2012cq} used during the MCMC, we employ the same uninformative flat priors for every parameter in each model and allow a sufficiently wide range for them.

Once the Bayesian Evidence is calculated, we define the Bayes Factor as the ratio of evidences between two models $M_{i}$ and $M_{j}$, namely $\mathcal{B}^{i}_{j} = \mathcal{E}_{i}/\mathcal{E}_{j}$: if $\mathcal{B}^{i}_{j} > 1$, $M_i$ is preferred over $M_j$, given the data (in this work, $\Lambda$CDM is the reference model $M_j$). For the sake of assessing model $M_i$ with respect to model $M_j$, we adopt Jeffreys' Scale \citep{Jeffreys:1939xee}: if $\ln \mathcal{B}^{i}_{j} < 1$, the evidence in favour of model $M_i$ is not significant; if $1 < \ln \mathcal{B}^{i}_{j} < 2.5$, it is substantial; if $2.5 < \ln \mathcal{B}^{i}_{j} < 5$, it is strong; if $\mathcal{B}^{i}_{j} > 5$, it is decisive. Negative values of $\ln \mathcal{B}^{i}_{j}$ can be easily interpreted as evidence against model $M_i$ (or in favour of model $M_j$).

\subsection{Type Ia Supernovae}
The Pantheon compilation \cite{Scolnic:2017caz} contains $1048$ objects within the redshift range $0.01<z<2.26$. We define the related $\chi^2_{SN}$ as
\begin{equation}
\chi^2_{SN} = \Delta \boldsymbol{\mathcal{\mu}}^{SN} \; \cdot \; \mathbf{C}^{-1}_{SN} \; \cdot \; \Delta  \boldsymbol{\mathcal{\mu}}^{SN} \;,
\end{equation}
where $\Delta\boldsymbol{\mathcal{\mu}} = \mathcal{\mu}_{\rm theo} - \mathcal{\mu}_{\rm obs}$ is the difference between the theoretical and the observed value of the distance modulus for each SNeIa and $\mathbf{C}_{SN}$ represents the total covariance matrix. The theoretically predicted distance modulus is defined as
\begin{equation}
\mu(z,\boldsymbol{p}) = 5 \log_{10} [ d_{L}(z, \boldsymbol{p}) ] +\mu_0 \; ,
\end{equation}
where
\begin{equation}
d_L(z,\boldsymbol{p})=(1+z)\int_{0}^{z}\frac{dz'}{E(z',\boldsymbol{p})} \,
\end{equation}
is the dimensionless luminosity distance and $\boldsymbol{p}$ is the vector of cosmological parameters. We also need to marginalize over the nuisance parameter $\mu_0$ (a combination of the Hubble constant, the speed
of light $c$ and the SNeIa absolute magnitude); following \cite{conley} we end up with:
\begin{equation}\label{eq:chis}
\chi^2_{SN}=a+\log \left(\frac{d}{2\pi}\right)-\frac{b^2}{d},
\end{equation}
where $a\equiv\left(\Delta \boldsymbol{\mathcal{\mu}}_{SN}\right)^T \; \cdot \; \mathbf{C}^{-1}_{SN} \; \cdot \; \Delta  \boldsymbol{\mathcal{\mu}}_{SN}$, $b\equiv\left(\Delta \boldsymbol{\mathcal{\mu}}^{SN}\right)^T \; \cdot \; \mathbf{C}^{-1}_{SN} \; \cdot \; \boldsymbol{1}$, $d\equiv\boldsymbol{1}\; \cdot \; \mathbf{C}^{-1}_{SN} \; \cdot \;\boldsymbol{1}$ and $\boldsymbol{1}$ is the identity matrix.

\subsection{Cosmic Chronometers}

The definition of Cosmic Chronometers (CC) applies to Early-Type Galaxies (ETGs) exhibiting a passive evolution, which provide measurements of the Hubble parameter over extended redshift ranges. The sample we use here covers the range $0<z<1.97$ \cite{moresco}. The $\chi^2_{H}$ in this case can be constructed as
\begin{equation}\label{eq:hubble_data}
\chi^2_{H}= \sum_{i=1}^{24} \frac{\left( H(z_{i},\boldsymbol{p})-H_{obs}(z_{i}) \right)^{2}}{\sigma^2_{H}(z_{i})} \; ,
\end{equation}
where $\sigma_{H}(z_{i})$ are the observational errors on the measured values $H_{obs}(z_{i})$.

\subsection{H0LiCOW}

The H0LiCOW collaboration \cite{Suyu:2016qxx} used the sensitivity of strong gravitational lensing events to constrain $H_0$ and the cosmological background. In particular, it focused on $6$ selected lensed quasars \cite{Wong:2019kwg} for which multiple images were provided. It is well-known that the light travel time from the quasars (sources) to the observer depends on the path length and the gravitational potential of the foreground mass (lens), so that multiple images can exhibit a time delay at collection given by
\begin{equation}\label{timedelGO}
    t(\boldsymbol{\theta},\boldsymbol{\beta})=\frac{1+z_L}{c}\frac{D_L D_S}{D_{LS}}\left[\frac{1}{2}(\boldsymbol{\theta}-\boldsymbol{\beta})^2-
    \hat{\Psi}(\boldsymbol{\theta})\right].
\end{equation}
As in a typical gravitational lensing configuration \cite{gralen.boo}, $z_L$ is the lens redshift, $\boldsymbol{\theta}$ the angular position of the image, $\boldsymbol{\beta}$ the angular position of the source and $\hat{\Psi}$ the effective lens potential, while $D_S$, $D_L$ and $D_{LS}$ are, respectively, the angular diameter distances from the source to the observer, from the lens to the observer, and between source and lens. They are given by the following expression:
\begin{equation}
D_{A}(z,\boldsymbol{p})=\frac{1}{1+z}\int_{0}^{z} \frac{c\, \mathrm{d}z'}{H(z',\boldsymbol{p})} \;
\end{equation}
with $D_{S} = D_{A}(z_S)$, $D_{L} = D_{A}(z_L)$, and $D_{LS} = 1/(1+z_S) \left[(1+z_S)D_S - (1+z_L)D_L\right]$ in the case of spatial flatness \cite{Hogg:1999ad}.

The quantity used in our analysis is generally called time-delay distance, and is defined as
\begin{equation}
D_{\Delta t} \equiv (1+z_L)\frac{D_L D_S}{D_{LS}}\, ;
\end{equation}
all data ($D^{obs}_{\Delta t,i}$) and errors $(\sigma_{D_{\Delta t,i}})$ on this quantity for each quasar are provided in \cite{Wong:2019kwg}. Thus, the $\chi^2$ for H0LiCOW data is
\begin{equation}\label{eq:cow_data}
\chi^2_{HCOW}= \sum_{i=1}^{6} \frac{\left( D_{\Delta t,i}(\boldsymbol{p})-D^{obs}_{\Delta t,i}\right)^{2}}{\sigma^2_{D_{\Delta t,i}}} \; ,
\end{equation}

\subsection{Gamma Ray Bursts}

The ``Mayflower'' sample is made of 79 GRBs within the redshift range $1.44<z<8.1$ \cite{Liu:2014vda}. Given that GRBs observable is again a distance modulus, the same procedure used in the section above for SNeIa also applies here; the final $\chi_{G}^2$ estimator is given by \eqref{eq:chis} as well,
with $a\equiv \left(\Delta\boldsymbol{\mathcal{\mu}}^{G}\right)^T \, \cdot \, \mathbf{C}^{-1}_{G} \, \cdot \, \Delta  \boldsymbol{\mathcal{\mu}}^{G}$, $b\equiv\left(\Delta \boldsymbol{\mathcal{\mu}}^{G}\right)^T \, \cdot \, \mathbf{C}^{-1}_{G} \, \cdot \, \boldsymbol{1}$ and $d\equiv\boldsymbol{1}\, \cdot \, \mathbf{C}^{-1}_{G} \, \cdot \, \boldsymbol{1}$.

\subsection{Baryon Acoustic Oscillations}

The $\chi^2$ estimator for BAO data is the sum of different contributions
\begin{equation}
\chi^2_{BAO} = \Delta \boldsymbol{\mathcal{F}}^{BAO} \, \cdot \ \mathbf{C}^{-1}_{BAO} \, \cdot \, \Delta  \boldsymbol{\mathcal{F}}^{BAO} \ ,
\end{equation}
where the observables $\mathcal{F}^{BAO}$ change according to the chosen survey. We employ data from the WiggleZ Dark Energy Survey (at redshifts $0.44$, $0.6$ and $0.73$) \cite{Blake:2012pj}, for which the relevant physical quantities are the acoustic parameter
\begin{equation}\label{eq:AWiggle}
A(z,\boldsymbol{p}) = 100  \sqrt{\Omega_{m} \, h^2} \frac{D_{V}(z,\boldsymbol{p})}{c \, z} \, ,
\end{equation}
and the Alcock-Paczynski distortion parameter
\begin{equation}\label{eq:FWiggle}
F(z,\boldsymbol{p}) = (1+z)  \frac{D_{A}(z,\boldsymbol{p})\, H(z,\boldsymbol{p})}{c} \, ,
\end{equation}
where
\begin{equation}
D_{A}(z,\boldsymbol{p})=\frac{1}{1+z}\int_{0}^{z} \frac{c\, \mathrm{d}z'}{H(z',\boldsymbol{p})} \;
\end{equation}
is the angular diameter distance and
\begin{equation}
D_{V}(z,\boldsymbol{p})=\left[ (1+z)^2 D^{2}_{A}(z,\boldsymbol{p}) \frac{c z}{H(z,\boldsymbol{p})}\right]^{1/3}
\end{equation}
is the volume distance, the geometric mean of the radial $(\propto H^{-1})$ and tangential $(D_A)$ BAO modes.

We consider multiple data from the SDSS-III Baryon Oscillation Spectroscopic Survey (BOSS). From the DR$12$ analysis in \cite{Alam:2016hwk}, the following quantities are given
\begin{equation}
D_{M}(z,\boldsymbol{p}) \frac{r^{fid}_{s}(z_{d},)}{r_{s}(z_{d},\boldsymbol{p})}, \qquad H(z) \frac{r_{s}(z_{d},\boldsymbol{p})}{r^{fid}_{s}(z_{d})} \,.
\end{equation}
where the comoving distance $D_M$ is
\begin{equation}
D_{M}(z,\boldsymbol{p})=\int_{0}^{z} \frac{c\, \mathrm{d}z'}{H(z',\boldsymbol{p})} \; .
\end{equation}
Here, $r_{s}(z_{d})$ denotes the sound horizon evaluated at the dragging redshift, while $r^{fid}_{s}(z_{d})$ is the sound horizon calculated at a given fiducial cosmological model (in this case, it is $147.78$ Mpc). The dragging redshift is estimated using the approximation \cite{Eisenstein:1997ik} as
\begin{equation}\label{eq:zdrag}
z_{d} = \frac{1291 (\Omega_{m} \, h^2)^{0.251}}{1+0.659(\Omega_{m} \, h^2)^{0.828}} \left[ 1+ b_{1} (\Omega_{b} \, h^2)^{b2}\right]\; ,
\end{equation}
where the factors $b_1$ and $b_2$ are given by
\begin{eqnarray}\label{eq:zdrag_b}
b_{1} &=& 0.313 (\Omega_{m} \, h^2)^{-0.419} \left[ 1+0.607 (\Omega_{m} \, h^2)^{0.6748}\right] \,,
	\nonumber \\
b_{2} &=& 0.238 (\Omega_{m} \, h^2)^{0.223}\,,
\end{eqnarray}
respectively. The sound horizon is defined as:
\begin{equation}\label{eq:soundhor}
r_{s}(z,\boldsymbol{p}) = \int^{\infty}_{z} \frac{c_{s}(z')}{H(z',\boldsymbol{p})} \mathrm{d}z'\, ,
\end{equation}
where the sound speed is given by
\begin{equation}\label{eq:soundspeed}
c_{s}(z) = \frac{c}{\sqrt{3(1+\overline{R}_{b}\, (1+z)^{-1})}} \; ,
\end{equation}
and the baryon-to-photon density ratio parameters is $\overline{R}_{b}= 31500 \Omega_{b} \, h^{2} \left( T_{CMB}/ 2.7 \right)^{-4}$, with $T_{CMB} = 2.726$ K.

From the DR$12$ we also include measurements derived from the void-galaxy cross-correlation \cite{Nadathur:2019mct}
\begin{align}
&\frac{D_{A}(z=0.57)}{r_{s}(z_{d})} = 9.383 \pm 0.077\,, \\
&H(z=0.57) r_{s}(z_{d})=(14.05 \pm 0.14)\, 10^{3} \rm{km/s}\,.
\end{align}

From the extended Baryon Oscillation Spectroscopic Survey (eBOSS) we have used the point $D_V(z=1.52)=3843\pm147\dfrac{r_s(zd)}{r_s^{fid}(z_d)}$ Mpc \cite{Ata:2017dya}.

Additional points taken into account from eBOSS DR14 are obtained from the combination of the quasar Lyman-$\alpha$ autocorrelation function \cite{lyman} with the cross-correlation measurement \cite{Blomqvist:2019rah}, namely
\begin{eqnarray}
\frac{D_{A}(z=2.34)}{r_{s}(z_{d})} &=& 36.98^{+1.26}_{-1.18} \\
\frac{c}{H(z=2.34) r_{s}(z_{d})}  &=& 9.00^{+0.22}_{-0.22}\, .
\end{eqnarray}

\subsection{Cosmic Microwave Background}
\label{cmb}

Regarding CMB data, we use the shift parameters \cite{Wang:2007mza} derived from the latest \textit{Planck} $2018$ data release \cite{cmb&sn}. In this case the $\chi^2_{CMB}$ is defined as
\begin{equation}
\chi^2_{CMB} = \Delta \boldsymbol{\mathcal{F}}^{CMB} \; \cdot \; \mathbf{C}^{-1}_{CMB} \; \cdot \; \Delta  \boldsymbol{\mathcal{F}}^{CMB} \; ,
\end{equation}
where the vector $\mathcal{F}^{CMB}$ contains the quantities
\begin{eqnarray}
R(\boldsymbol{p}) &\equiv& \sqrt{\Omega_m H^2_{0}} \frac{r(z_{\ast},\boldsymbol{p})}{c} \nonumber \\
l_{a}(\boldsymbol{p}) &\equiv& \pi \frac{r(z_{\ast},\boldsymbol{p})}{r_{s}(z_{\ast},\boldsymbol{p})}\,,
\end{eqnarray}
in addition to $\Omega_b\,h^2$. Here, $r_{s}(z_{\ast})$ is the comoving sound horizon evaluated at the photon-decoupling redshift given by \cite{Hu:1995en} by the following fitting formula,
\begin{eqnarray}{\label{eq:zdecoupl}}
z_{\ast} &=& 1048 \left[ 1 + 0.00124 (\Omega_{b} h^{2})^{-0.738}\right] \cdot   \nonumber \\
&& \cdot \left(1+g_{1} (\Omega_{m} h^{2})^{g_{2}} \right)  \,,
\end{eqnarray}
where the factors $g_1$ and $g_2$ are
\begin{eqnarray}
g_{1} &=& \frac{0.0783 (\Omega_{b} h^{2})^{-0.238}}{1+39.5(\Omega_{b} h^{2})^{-0.763}} \nonumber \\
g_{2} &=& \frac{0.560}{1+21.1(\Omega_{b} h^{2})^{1.81}} \,, \nonumber
\end{eqnarray}
while $r$ is the comoving distance defined by
\begin{equation}
r(z,\boldsymbol{p})  = \int_{0}^{z} \frac{c\, \mathrm{d}z'}{H(z',\boldsymbol{p})} \; .
\end{equation}

\section{Results and Discussion}
\label{results}

{\renewcommand{\tabcolsep}{1.5mm}
{\renewcommand{\arraystretch}{2.}
\begin{table*}
\begin{minipage}{0.75\textwidth}
\caption{Results of our analysis at 1$\sigma$ confidence level.}\label{tab:results}
\centering
\resizebox*{\textwidth}{!}{
\begin{tabular}{c|cc|cc|c}
\hline
   &  \multicolumn{2}{c}{$\Lambda{\rm CDM}$} & \multicolumn{2}{c}{Quadratic GUP} & Linear GUP \\
\hline
   & late  &  full &  late & full & late \\
\hline
\hline
$\Omega_{ m}$   & $0.292^{+0.017}_{-0.016}$ & $0.319^{+0.005}_{-0.005}$    & $0.296^{+0.018}_{-0.017}$ & $0.320^{+0.005}_{-0.005}$  & $0.369^{+0.023}_{-0.021}$ \\
$\Omega_{ b}$   & $-$                       & $0.0494^{+0.0004}_{-0.0004}$ & $-$                        & $0.0495^{+0.0004}_{-0.0004}$ & $-$ \\
$h$                & $0.712^{+0.013}_{-0.013}$ & $0.673^{+0.003}_{-0.003}$    & $0.713^{+0.013}_{-0.012}$  & $0.672^{+0.003}_{-0.004}$ & $0.732^{+0.014}_{-0.013}$ \\
$\log \left(\dfrac{\alphaq \lp}{c}\right)$ & $-$ & $-$  & $<-2.29$ & $<-12.52$ & $-1.841^{+0.020}_{-0.013}$ \\
\hline
$\Omega_{\Lambda}$ & $0.707^{+0.016}_{-0.017}$ & $0.680^{+0.005}_{-0.005}$    & $0.708^{+0.017}_{-0.016}$  & $0.680^{+0.005}_{-0.005}$ & $0.978^{+0.024}_{-0.025}$ \\
$\alphaq^2$        & $-$                       & $-$                          & $<9.31 \cdot 10^{81}$      & $<5.16 \cdot 10^{59}$ & $\left(2.78^{+0.26}_{-0.17}\right)\cdot 10^{83}$ \\
$\alphaq^2 \lp^2$  & $-$                       & $-$                          & $<2.38 \cdot 10^{12}$      & $<1.34 \cdot 10^{-10}$ & $\left(7.22^{+0.68}_{-0.43}\right)\cdot 10^{13}$ \\
$\chi^2$                        & $1094.17$      & $1124.23$      & $1093.77$                  & $1124.25$ & $1121.12$ \\
$\mathcal{B}^{i}_{j}$           & $\mathit{1}$   & $\mathit{1}$   & $1.05^{+0.03}_{-0.03}$     & $0.92^{+0.03}_{-0.03}$  & $(3.65^{+0.11}_{-0.09})\cdot10^{-6}$ \\
$\log \mathcal{B}^{i}_{j}$      & $\mathit{0}$   & $\mathit{0}$   & $0.05^{+0.02}_{-0.03}$     & $-0.09^{+0.03}_{-0.03}$ & $-12.52^{+0.03}_{-0.03}$ \\
\hline
\hline
\end{tabular}}
\end{minipage}
\end{table*}}}

The results of our analysis are summarised in Table \ref{tab:results} and complemented by the comparison with a reference model, namely the standard $\Lambda$CDM. In the upper part of the table, we show the primary cosmological parameters as constrained by our statistical analysis, i.e.\ $\Omega_{ m}$, $\Omega_{ b}$ and $h$, with the addition of $\log \left(\alphaq \lp / c\right)$, where the $\log$ is used for ease of numerical computation. In the lower part of the table, we show some secondary parameters which can be derived from the primary ones and include the constraints on the GUP parameters that are central to our study.
More specifically, we report
\begin{itemize}
 \item $\Omega_{\Lambda}$, the cosmological constant (dark energy) contribution derived from the normalization condition $E(z=0)=1$. It allows us to explore the possibility that the GUP correction could play the role of an effective dark energy fluid;
 \item $\alphaq^2$ or $\alphal^2\sim \beta$ (for the quadratic and linear formulations of the GUP, respectively). They represent the quantities constrained by experiments and reported in Table \ref{tab:bounds};
 \item $\alphaq^2 \lp^2$, the parameter we perturbatively expand around to find $f_{G}(A)$ in \eqref{fga}, the departure function from the HUP that underlies this entire study. Thus, \textit{a posteriori} verifying the smallness of its value represents an important consistency check for our analysis and a way to assess the reliability of our conclusions;
 \item $\chi^{2}_{\rm min}$, $\mathcal{B}^{i}_{j}$ and $\log \mathcal{B}^{i}_{j}$, which are the minimum value of $\chi^2$, the Bayes Factor and its logarithm (for comparison with Jeffreys' scale), respectively.
\end{itemize}

\subsection{Quadratic GUP}

The most conclusive case in our analysis is that of the quadratic GUP with the full data set. As shown by the Bayesian Factor in Table \ref{tab:results}, this model is essentially indistinguishable from the standard $\Lambda$CDM case and the evidence in its favour is negligible. This means, as expected, that the GUP correction on cosmological scales plays a negligible role and therefore it cannot account for a dark energy fluid at all. This claim can be additionally supported if we consider \eqref{ez}: eliminating the dark energy parameter and solving the normalization condition $E(z=0)=1$ for $\alphaq \lp / c$ yields
\begin{equation}
  \frac{\alphaq \lp}{c} = \sqrt{\frac{32(\Omega_k+\Omega_m+\Omega_r-1)}{H^{2}_0 (-1+\Omega_{ k})^2}},
\end{equation}
which entails
\begin{equation}
\Omega_{k}+\Omega_{m}+\Omega_{r}>1 \, .
\end{equation}
Given the best present estimates of $\Omega_{m}$ and $\Omega_{r}$, this condition would be satisfied by values $\Omega_{k}\gtrsim 0.68$. However, such value appears to be incompatible not only with the most stringent available constraint on the spatial curvature of our universe (obtained from \textit{Planck}, see e.g.\ \cite{Aghanim:2018eyx}), but also with the recent claim, still supported by \textit{Planck} data, of a positively curved (i.e.\ $\Omega_{k}<0$) universe \cite{DiValentino:2019qzk, handley}.

The impossibility of having the GUP corrections account for dark energy is also in agreement with the discussion provided in \cite{kouwn}, where it was shown that the GUP effects would amount to a dark energy density $\sim\alpha H^4$. This scaling would cause them to decrease extremely quickly, meaning that the GUP alone would not be able to explain the current acceleration of the Universe at all. Other authors note that, even if the energy density were of the order $\sim H^2$, it would still not be sufficient to account for the present acceleration \cite{maggiorede}. A different perspective is provided by \cite{paliathanasis}.

The main goal of our analysis is to find an estimate for the GUP parameter, compatibly with the accuracy of the cosmological probes we used. The result obtained for the quadratic GUP model with the full data set is approximately $\alphaq^2<10^{59}$. This bound is clearly much less stringent than most of those obtained with quantum experiments, but, to the best of our knowledge, it represents one of the tightest cosmological and astrophysical constraints achieved so far, as shown in Table \ref{tab:bounds}. Moreover, it is easy to see from Table \ref{tab:results} that $\alphaq^2 \lp^2$ is sufficiently small, being $\alphaq^2 \lp^2<10^{-10}$, which ensures the appropriateness of the Taylor expansion in \eqref{fga}.

However, when considering late-time observations only, the picture changes quite considerably. Once again, the GUP-influenced model is statistically equivalent to the corresponding $\Lambda$CDM, which provides further support for the claim that the GUP contributions cannot account for dark energy. Indeed, we have considered late-time observations precisely for this reason: on the one hand, it is well-known that, although the CMB geometrical data are very precise if compared to other probes, they are also intrinsically biased towards a cosmological constant model for dark energy. On the other hand, if any dynamical dark energy behaviour could appear, it would be more manifest for $z<2$, a range which is fully covered by the late-time data we employed. In conclusion, the full equivalence of $\Lambda$CDM and the quadratic GUP model at late times clearly indicates a negligible role of the GUP on cosmological scales.

The bound on the GUP parameter obtained with late-time data is much more relaxed than that obtained with the full data set: it is $\alphaq^2<10^{81}$, namely one of the least stringent constraint in Table \ref{tab:bounds} (although it perfectly agrees with that found in \cite{kouwn}). It is also worth noting that $\alphaq^2 \lp^2<10^{12}$, much larger than that obtained with the full data set as well. This means, \textit{a posteriori}, that any conclusion drawn from the late-time case should be handled carefully: although we are only able to find an upper bound on $\alphaq \lp$ (meaning that part of the parameter space explored by the MCMC would still involve small values of this quantity) it should also be noted that all expressions containing the departure function \eqref{fga} have been partly evaluated outside their region of validity. This technically represents an extrapolation; therefore, the results are not on a footing as strong as those in the analysis employing the full data set.

\subsection{Linear GUP}

Testing the Friedmann equations modified by the linear GUP has proved way more challenging than in the quadratic GUP case, which also affects the reliability of the conclusions we can draw. We remark that our analysis has focused on the case $\alphal>0$ only, as clarified in Section \ref{lineargup}. First and foremost, it is important to point out the impossibility of fitting early-time data (i.e.\ CMB and BAO data from BOSS/eBOSS) with a linear GUP model. Performing an extensive exploration of parameter space, we found that the $\chi^2$ could not be lower than $\sim 10^7$, indicating a striking inconsistency between data and model.

Consequently, we tried to check if such inconsistency could be relaxed by including late-time data only, since these naturally allow for more freedom in the dark sector. We were able to obtain a much better fit, but also new problems arose. To begin with, the value of the Bayesian Factor is very small, leading to striking decisive evidence against the linear GUP model.

Concerning the values of the cosmological parameters, we found $\Omega_{m}\approx 0.37$, which is much larger than expected in the standard lore. The associated errors were quite large as well, but still a $\sim 2\sigma$ tension was present with the standard value.
Additionally, also the cosmological constant contribution is found to be large, namely $\Omega_{\Lambda}\approx 0.98$, which would appear to suggest a strong decelerating effect of the linear GUP contribution.

The constraint on the GUP parameter is $\alphal^2 \lesssim 10^{83}$. The weakness of this bound is expected, especially given the lack of precision of late-time cosmological data.
However, we remark that all the previous results should be taken with care, because the value of the parameter $\alpha^2\lp^2\approx 10^{13}$ is also quite large and comparable to the case of the quadratic GUP model with late-time data.

To summarize, the previous analysis shows that the linear GUP model strongly disagrees with the presently available cosmological data, which greatly limits the amount of relevant information we can extract from comparison between them.

\section{Conclusions}

An abundance of cosmological models often motivated by advances in high-energy physics has been devised in the past decades, including some endowed with a GUP. However, not all of these models have been put through observational tests (especially those dealing with a GUP) even now that the technology to make a wealth of high-precision observations is available. In this paper, we provided an example of testing two such models, whose physical predictions ended up being starkly different.

More specifically, we studied two models arising from Friedmann equations influenced by the GUP, in two formulations involving, respectively, a linear and a quadratic term in momentum uncertainty. We put their cosmological viability to the test by means of a statistical analysis relying on well-known cosmological probes at early and late times. On the one hand, we found that the model endowed with the quadratic GUP mimics $\Lambda$CDM very closely when the full set of early and late probes is employed, which means that the GUP contribution cannot play the role of an alternative dark energy fluid. However, the main result of this work is the interesting constraint on the GUP parameter, namely $\alphaq^2<10^{59}$, which, although not as stringent as those coming from quantum experiments, is tighter than most of those obtained from gravitational measurements and is one of the few obtained with cosmological data. On the other hand, the model arising from the linear GUP does not appear to be compatible with the cosmological data we analysed. This shows that not all formulations are equally suitable to find constraints in a specific physical system, since, for example, the linear formulation was found more suitable when using gravitational waves \cite{gw}.

In the future, it would be compelling to study other cosmological models influenced by other formulations of the GUP. Although the low-energy window for quantum gravity phenomenology, far from Planck scale, is at present only able to constrain the GUP parameter and not measure it exactly, the ever-improving quality of these constraints appears very promising. This work also illustrates the potential held by high-precision cosmological data to constrain the GUP parameter, thus complementing the bounds obtained from gravitational measurements and quantum experiments.

\section*{Acknowledgements}
The authors would like to thank Mariusz P. D\c{a}browski, Roberto Casadio and Hussain Gohar for useful comments and discussions. S.G. was supported by the project ``Uniwersytet 2.0 - Strefa Kariery, ``Miedzynarodowe Studia Doktoranckie Nauk Scislych WMF'', nr POWR.03.05.00-00-Z064/17-00''.

\bibliographystyle{apsrev4-1}

\begin{thebibliography}{84}%
\makeatletter
\providecommand \@ifxundefined [1]{%
 \@ifx{#1\undefined}
}%
\providecommand \@ifnum [1]{%
 \ifnum #1\expandafter \@firstoftwo
 \else \expandafter \@secondoftwo
 \fi
}%
\providecommand \@ifx [1]{%
 \ifx #1\expandafter \@firstoftwo
 \else \expandafter \@secondoftwo
 \fi
}%
\providecommand \natexlab [1]{#1}%
\providecommand \enquote  [1]{``#1''}%
\providecommand \bibnamefont  [1]{#1}%
\providecommand \bibfnamefont [1]{#1}%
\providecommand \citenamefont [1]{#1}%
\providecommand \href@noop [0]{\@secondoftwo}%
\providecommand \href [0]{\begingroup \@sanitize@url \@href}%
\providecommand \@href[1]{\@@startlink{#1}\@@href}%
\providecommand \@@href[1]{\endgroup#1\@@endlink}%
\providecommand \@sanitize@url [0]{\catcode `\\12\catcode `\$12\catcode
  `\&12\catcode `\#12\catcode `\^12\catcode `\_12\catcode `\%12\relax}%
\providecommand \@@startlink[1]{}%
\providecommand \@@endlink[0]{}%
\providecommand \url  [0]{\begingroup\@sanitize@url \@url }%
\providecommand \@url [1]{\endgroup\@href {#1}{\urlprefix }}%
\providecommand \urlprefix  [0]{URL }%
\providecommand \Eprint [0]{\href }%
\providecommand \doibase [0]{http://dx.doi.org/}%
\providecommand \selectlanguage [0]{\@gobble}%
\providecommand \bibinfo  [0]{\@secondoftwo}%
\providecommand \bibfield  [0]{\@secondoftwo}%
\providecommand \translation [1]{[#1]}%
\providecommand \BibitemOpen [0]{}%
\providecommand \bibitemStop [0]{}%
\providecommand \bibitemNoStop [0]{.\EOS\space}%
\providecommand \EOS [0]{\spacefactor3000\relax}%
\providecommand \BibitemShut  [1]{\csname bibitem#1\endcsname}%
\let\auto@bib@innerbib\@empty
\bibitem [{\citenamefont {Hossenfelder}(2013)}]{hossenfelder}%
  \BibitemOpen
  \bibfield  {author} {\bibinfo {author} {\bibfnamefont {S.}~\bibnamefont
  {Hossenfelder}},\ }\href {\doibase 10.12942/lrr-2013-2} {\bibfield  {journal}
  {\bibinfo  {journal} {Living Rev. Rel.}\ }\textbf {\bibinfo {volume} {16}},\
  \bibinfo {pages} {2} (\bibinfo {year} {2013})},\ \Eprint
  {http://arxiv.org/abs/1203.6191} {arXiv:1203.6191} \BibitemShut {NoStop}%
\bibitem [{\citenamefont {Mead}(1964)}]{mead}%
  \BibitemOpen
  \bibfield  {author} {\bibinfo {author} {\bibfnamefont {C.}~\bibnamefont
  {Mead}},\ }\href {\doibase 10.1103/PhysRev.135.B849} {\bibfield  {journal}
  {\bibinfo  {journal} {Phys.\ Rev.}\ }\textbf {\bibinfo {volume} {135}},\
  \bibinfo {pages} {B849} (\bibinfo {year} {1964})}\BibitemShut {NoStop}%
\bibitem [{\citenamefont {Hawking}(1975)}]{hawking}%
  \BibitemOpen
  \bibfield  {author} {\bibinfo {author} {\bibfnamefont {S.}~\bibnamefont
  {Hawking}},\ }\href {\doibase 10.1007/BF02345020} {\bibfield  {journal}
  {\bibinfo  {journal} {Commun.\ Math.\ Phys.}\ }\textbf {\bibinfo {volume}
  {43}},\ \bibinfo {pages} {199} (\bibinfo {year} {1975})},\ \bibinfo {note}
  {[Erratum: Commun.Math.Phys. 46, 206 (1976)]}\BibitemShut {NoStop}%
\bibitem [{\citenamefont {Martin}\ and\ \citenamefont
  {Brandenberger}(2001)}]{brandenberger}%
  \BibitemOpen
  \bibfield  {author} {\bibinfo {author} {\bibfnamefont {J.}~\bibnamefont
  {Martin}}\ and\ \bibinfo {author} {\bibfnamefont {R.~H.}\ \bibnamefont
  {Brandenberger}},\ }\href {\doibase 10.1103/PhysRevD.63.123501} {\bibfield
  {journal} {\bibinfo  {journal} {Phys. Rev. D}\ }\textbf {\bibinfo {volume}
  {63}},\ \bibinfo {pages} {123501} (\bibinfo {year} {2001})},\ \Eprint
  {http://arxiv.org/abs/hep-th/0005209} {arXiv:hep-th/0005209} \BibitemShut
  {NoStop}%
\bibitem [{\citenamefont {Amati}\ \emph {et~al.}(1989)\citenamefont {Amati},
  \citenamefont {Ciafaloni},\ and\ \citenamefont {Veneziano}}]{amati}%
  \BibitemOpen
  \bibfield  {author} {\bibinfo {author} {\bibfnamefont {D.}~\bibnamefont
  {Amati}}, \bibinfo {author} {\bibfnamefont {M.}~\bibnamefont {Ciafaloni}}, \
  and\ \bibinfo {author} {\bibfnamefont {G.}~\bibnamefont {Veneziano}},\ }\href
  {\doibase 10.1016/0370-2693(89)91366-X} {\bibfield  {journal} {\bibinfo
  {journal} {Phys. Lett. B}\ }\textbf {\bibinfo {volume} {216}},\ \bibinfo
  {pages} {41} (\bibinfo {year} {1989})}\BibitemShut {NoStop}%
\bibitem [{\citenamefont {Veneziano}(1989)}]{veneziano}%
  \BibitemOpen
  \bibfield  {author} {\bibinfo {author} {\bibfnamefont {G.}~\bibnamefont
  {Veneziano}},\ }\href {http://cds.cern.ch/record/197729} {\bibfield
  {journal} {\bibinfo  {journal} {Conf. Proc. C}\ }\textbf {\bibinfo {volume}
  {8903131}},\ \bibinfo {pages} {86} (\bibinfo {year} {1989})}\BibitemShut
  {NoStop}%
\bibitem [{\citenamefont {Adler}\ and\ \citenamefont {Santiago}(1999)}]{adler}%
  \BibitemOpen
  \bibfield  {author} {\bibinfo {author} {\bibfnamefont {R.~J.}\ \bibnamefont
  {Adler}}\ and\ \bibinfo {author} {\bibfnamefont {D.~I.}\ \bibnamefont
  {Santiago}},\ }\href {\doibase 10.1142/S0217732399001462} {\bibfield
  {journal} {\bibinfo  {journal} {Mod. Phys. Lett. A}\ }\textbf {\bibinfo
  {volume} {14}},\ \bibinfo {pages} {1371} (\bibinfo {year} {1999})},\ \Eprint
  {http://arxiv.org/abs/gr-qc/9904026} {arXiv:gr-qc/9904026} \BibitemShut
  {NoStop}%
\bibitem [{\citenamefont {Maggiore}(1993)}]{maggiore}%
  \BibitemOpen
  \bibfield  {author} {\bibinfo {author} {\bibfnamefont {M.}~\bibnamefont
  {Maggiore}},\ }\href {\doibase 10.1016/0370-2693(93)91401-8} {\bibfield
  {journal} {\bibinfo  {journal} {Phys. Lett. B}\ }\textbf {\bibinfo {volume}
  {304}},\ \bibinfo {pages} {65} (\bibinfo {year} {1993})},\ \Eprint
  {http://arxiv.org/abs/hep-th/9301067} {arXiv:hep-th/9301067} \BibitemShut
  {NoStop}%
\bibitem [{\citenamefont {Bronstein}(1936)}]{bronstein}%
  \BibitemOpen
  \bibfield  {author} {\bibinfo {author} {\bibfnamefont {M.}~\bibnamefont
  {Bronstein}},\ }\href@noop {} {\bibfield  {journal} {\bibinfo  {journal}
  {Phys. Z. Sowjetunion}\ }\textbf {\bibinfo {volume} {9}},\ \bibinfo {pages}
  {140} (\bibinfo {year} {1936})}\BibitemShut {NoStop}%
\bibitem [{\citenamefont {Gorelik}(2005)}]{gorelik}%
  \BibitemOpen
  \bibfield  {author} {\bibinfo {author} {\bibfnamefont {G.~E.}\ \bibnamefont
  {Gorelik}},\ }\href {\doibase 10.1070/pu2005v048n10abeh005820} {\bibfield
  {journal} {\bibinfo  {journal} {Physics-Uspekhi}\ }\textbf {\bibinfo {volume}
  {48}},\ \bibinfo {pages} {1039} (\bibinfo {year} {2005})}\BibitemShut
  {NoStop}%
\bibitem [{\citenamefont {Scardigli}(1999)}]{scardigli}%
  \BibitemOpen
  \bibfield  {author} {\bibinfo {author} {\bibfnamefont {F.}~\bibnamefont
  {Scardigli}},\ }\href {\doibase 10.1016/S0370-2693(99)00167-7} {\bibfield
  {journal} {\bibinfo  {journal} {Phys. Lett. B}\ }\textbf {\bibinfo {volume}
  {452}},\ \bibinfo {pages} {39} (\bibinfo {year} {1999})},\ \Eprint
  {http://arxiv.org/abs/hep-th/9904025v1} {arXiv:hep-th/9904025v1} \BibitemShut
  {NoStop}%
\bibitem [{\citenamefont {Ali}\ \emph {et~al.}(2009)\citenamefont {Ali},
  \citenamefont {Das},\ and\ \citenamefont {Vagenas}}]{discrete}%
  \BibitemOpen
  \bibfield  {author} {\bibinfo {author} {\bibfnamefont {A.~F.}\ \bibnamefont
  {Ali}}, \bibinfo {author} {\bibfnamefont {S.}~\bibnamefont {Das}}, \ and\
  \bibinfo {author} {\bibfnamefont {E.~C.}\ \bibnamefont {Vagenas}},\ }\href
  {\doibase 10.1016/j.physletb.2009.06.061} {\bibfield  {journal} {\bibinfo
  {journal} {Phys. Lett. B}\ }\textbf {\bibinfo {volume} {678}},\ \bibinfo
  {pages} {497} (\bibinfo {year} {2009})},\ \Eprint
  {http://arxiv.org/abs/0906.5396} {arXiv:0906.5396} \BibitemShut {NoStop}%
\bibitem [{\citenamefont {Alonso-Serrano}\ \emph {et~al.}(2018)\citenamefont
  {Alonso-Serrano}, \citenamefont {Dabrowski},\ and\ \citenamefont
  {Gohar}}]{anamariu}%
  \BibitemOpen
  \bibfield  {author} {\bibinfo {author} {\bibfnamefont {A.}~\bibnamefont
  {Alonso-Serrano}}, \bibinfo {author} {\bibfnamefont {M.~P.}\ \bibnamefont
  {Dabrowski}}, \ and\ \bibinfo {author} {\bibfnamefont {H.}~\bibnamefont
  {Gohar}},\ }\href {\doibase 10.1103/PhysRevD.97.044029} {\bibfield  {journal}
  {\bibinfo  {journal} {Phys. Rev.}\ }\textbf {\bibinfo {volume} {D97}},\
  \bibinfo {pages} {044029} (\bibinfo {year} {2018})},\ \Eprint
  {http://arxiv.org/abs/1801.09660} {arXiv:1801.09660} \BibitemShut {NoStop}%
\bibitem [{\citenamefont {Scardigli}\ \emph {et~al.}(2017)\citenamefont
  {Scardigli}, \citenamefont {Lambiase},\ and\ \citenamefont
  {Vagenas}}]{Scardigli:2016pjs}%
  \BibitemOpen
  \bibfield  {author} {\bibinfo {author} {\bibfnamefont {F.}~\bibnamefont
  {Scardigli}}, \bibinfo {author} {\bibfnamefont {G.}~\bibnamefont {Lambiase}},
  \ and\ \bibinfo {author} {\bibfnamefont {E.}~\bibnamefont {Vagenas}},\ }\href
  {\doibase 10.1016/j.physletb.2017.01.054} {\bibfield  {journal} {\bibinfo
  {journal} {Phys. Lett. B}\ }\textbf {\bibinfo {volume} {767}},\ \bibinfo
  {pages} {242} (\bibinfo {year} {2017})},\ \Eprint
  {http://arxiv.org/abs/1611.01469} {arXiv:1611.01469} \BibitemShut {NoStop}%
\bibitem [{\citenamefont {Adler}\ \emph {et~al.}(2001)\citenamefont {Adler},
  \citenamefont {Chen},\ and\ \citenamefont {Santiago}}]{adlerchen}%
  \BibitemOpen
  \bibfield  {author} {\bibinfo {author} {\bibfnamefont {R.~J.}\ \bibnamefont
  {Adler}}, \bibinfo {author} {\bibfnamefont {P.}~\bibnamefont {Chen}}, \ and\
  \bibinfo {author} {\bibfnamefont {D.~I.}\ \bibnamefont {Santiago}},\ }\href
  {\doibase 10.1023/A:1015281430411} {\bibfield  {journal} {\bibinfo  {journal}
  {Gen. Rel. Grav.}\ }\textbf {\bibinfo {volume} {33}},\ \bibinfo {pages}
  {2101} (\bibinfo {year} {2001})},\ \Eprint
  {http://arxiv.org/abs/gr-qc/0106080} {arXiv:gr-qc/0106080} \BibitemShut
  {NoStop}%
\bibitem [{\citenamefont {Casadio}\ and\ \citenamefont
  {Scardigli}(2014)}]{scardcasa}%
  \BibitemOpen
  \bibfield  {author} {\bibinfo {author} {\bibfnamefont {R.}~\bibnamefont
  {Casadio}}\ and\ \bibinfo {author} {\bibfnamefont {F.}~\bibnamefont
  {Scardigli}},\ }\href {\doibase 10.1140/epjc/s10052-013-2685-2} {\bibfield
  {journal} {\bibinfo  {journal} {Eur. Phys. J. C}\ }\textbf {\bibinfo {volume}
  {74}},\ \bibinfo {pages} {2685} (\bibinfo {year} {2014})},\ \Eprint
  {http://arxiv.org/abs/1306.5298} {arXiv:1306.5298} \BibitemShut {NoStop}%
\bibitem [{\citenamefont {Bekenstein}(1972)}]{Bekenstein:1972tm}%
  \BibitemOpen
  \bibfield  {author} {\bibinfo {author} {\bibfnamefont {J.}~\bibnamefont
  {Bekenstein}},\ }\href {\doibase 10.1007/BF02757029} {\bibfield  {journal}
  {\bibinfo  {journal} {Lett. Nuovo Cim.}\ }\textbf {\bibinfo {volume} {4}},\
  \bibinfo {pages} {737} (\bibinfo {year} {1972})}\BibitemShut {NoStop}%
\bibitem [{\citenamefont {Bekenstein}(1974)}]{Bekenstein:1974ax}%
  \BibitemOpen
  \bibfield  {author} {\bibinfo {author} {\bibfnamefont {J.~D.}\ \bibnamefont
  {Bekenstein}},\ }\href {\doibase 10.1103/PhysRevD.9.3292} {\bibfield
  {journal} {\bibinfo  {journal} {Phys. Rev. D}\ }\textbf {\bibinfo {volume}
  {9}},\ \bibinfo {pages} {3292} (\bibinfo {year} {1974})}\BibitemShut
  {NoStop}%
\bibitem [{\citenamefont {Hawking}(1974)}]{Hawking:1974rv}%
  \BibitemOpen
  \bibfield  {author} {\bibinfo {author} {\bibfnamefont {S.}~\bibnamefont
  {Hawking}},\ }\href {\doibase 10.1038/248030a0} {\bibfield  {journal}
  {\bibinfo  {journal} {Nature}\ }\textbf {\bibinfo {volume} {248}},\ \bibinfo
  {pages} {30} (\bibinfo {year} {1974})}\BibitemShut {NoStop}%
\bibitem [{\citenamefont {Medved}\ and\ \citenamefont
  {Vagenas}(2004)}]{vagenas}%
  \BibitemOpen
  \bibfield  {author} {\bibinfo {author} {\bibfnamefont {A.~J.~M.}\
  \bibnamefont {Medved}}\ and\ \bibinfo {author} {\bibfnamefont {E.~C.}\
  \bibnamefont {Vagenas}},\ }\href {\doibase 10.1103/PhysRevD.70.124021}
  {\bibfield  {journal} {\bibinfo  {journal} {Phys. Rev.}\ }\textbf {\bibinfo
  {volume} {D70}},\ \bibinfo {pages} {124021} (\bibinfo {year} {2004})},\
  \Eprint {http://arxiv.org/abs/hep-th/0411022} {arXiv:hep-th/0411022}
  \BibitemShut {NoStop}%
\bibitem [{\citenamefont {Anacleto}\ \emph {et~al.}(2015)\citenamefont
  {Anacleto}, \citenamefont {Brito},\ and\ \citenamefont {Passos}}]{anacleto}%
  \BibitemOpen
  \bibfield  {author} {\bibinfo {author} {\bibfnamefont {M.}~\bibnamefont
  {Anacleto}}, \bibinfo {author} {\bibfnamefont {F.}~\bibnamefont {Brito}}, \
  and\ \bibinfo {author} {\bibfnamefont {E.}~\bibnamefont {Passos}},\ }\href
  {\doibase 10.1016/j.physletb.2015.07.072} {\bibfield  {journal} {\bibinfo
  {journal} {Phys. Lett. B}\ }\textbf {\bibinfo {volume} {749}},\ \bibinfo
  {pages} {181} (\bibinfo {year} {2015})},\ \Eprint
  {http://arxiv.org/abs/1504.06295} {arXiv:1504.06295} \BibitemShut {NoStop}%
\bibitem [{\citenamefont {Jacobson}(1995)}]{jacobson}%
  \BibitemOpen
  \bibfield  {author} {\bibinfo {author} {\bibfnamefont {T.}~\bibnamefont
  {Jacobson}},\ }\href {\doibase 10.1103/PhysRevLett.75.1260} {\bibfield
  {journal} {\bibinfo  {journal} {Phys. Rev. Lett.}\ }\textbf {\bibinfo
  {volume} {75}},\ \bibinfo {pages} {1260} (\bibinfo {year} {1995})},\ \Eprint
  {http://arxiv.org/abs/gr-qc/9504004} {arXiv:gr-qc/9504004} \BibitemShut
  {NoStop}%
\bibitem [{\citenamefont {Cai}\ and\ \citenamefont {Kim}(2005)}]{caikim}%
  \BibitemOpen
  \bibfield  {author} {\bibinfo {author} {\bibfnamefont {R.-G.}\ \bibnamefont
  {Cai}}\ and\ \bibinfo {author} {\bibfnamefont {S.~P.}\ \bibnamefont {Kim}},\
  }\href {\doibase 10.1088/1126-6708/2005/02/050} {\bibfield  {journal}
  {\bibinfo  {journal} {JHEP}\ }\textbf {\bibinfo {volume} {02}},\ \bibinfo
  {pages} {050} (\bibinfo {year} {2005})},\ \Eprint
  {http://arxiv.org/abs/hep-th/0501055} {arXiv:hep-th/0501055} \BibitemShut
  {NoStop}%
\bibitem [{\citenamefont {Cai}\ \emph {et~al.}(2008)\citenamefont {Cai},
  \citenamefont {Cao},\ and\ \citenamefont {Hu}}]{caicao}%
  \BibitemOpen
  \bibfield  {author} {\bibinfo {author} {\bibfnamefont {R.-G.}\ \bibnamefont
  {Cai}}, \bibinfo {author} {\bibfnamefont {L.-M.}\ \bibnamefont {Cao}}, \ and\
  \bibinfo {author} {\bibfnamefont {Y.-P.}\ \bibnamefont {Hu}},\ }\href
  {\doibase 10.1088/1126-6708/2008/08/090} {\bibfield  {journal} {\bibinfo
  {journal} {JHEP}\ }\textbf {\bibinfo {volume} {08}},\ \bibinfo {pages} {090}
  (\bibinfo {year} {2008})},\ \Eprint {http://arxiv.org/abs/0807.1232}
  {arXiv:0807.1232} \BibitemShut {NoStop}%
\bibitem [{\citenamefont {Zhu}\ \emph {et~al.}(2009)\citenamefont {Zhu},
  \citenamefont {Ren},\ and\ \citenamefont {Li}}]{renli}%
  \BibitemOpen
  \bibfield  {author} {\bibinfo {author} {\bibfnamefont {T.}~\bibnamefont
  {Zhu}}, \bibinfo {author} {\bibfnamefont {J.-R.}\ \bibnamefont {Ren}}, \ and\
  \bibinfo {author} {\bibfnamefont {M.-F.}\ \bibnamefont {Li}},\ }\href
  {\doibase 10.1016/j.physletb.2009.03.020} {\bibfield  {journal} {\bibinfo
  {journal} {Phys. Lett. B}\ }\textbf {\bibinfo {volume} {674}},\ \bibinfo
  {pages} {204} (\bibinfo {year} {2009})},\ \Eprint
  {http://arxiv.org/abs/0811.0212} {arXiv:0811.0212} \BibitemShut {NoStop}%
\bibitem [{\citenamefont {Kouwn}(2018)}]{kouwn}%
  \BibitemOpen
  \bibfield  {author} {\bibinfo {author} {\bibfnamefont {S.}~\bibnamefont
  {Kouwn}},\ }\href {\doibase 10.1016/j.dark.2018.07.001} {\bibfield  {journal}
  {\bibinfo  {journal} {Phys. Dark Univ.}\ }\textbf {\bibinfo {volume} {21}},\
  \bibinfo {pages} {76} (\bibinfo {year} {2018})},\ \Eprint
  {http://arxiv.org/abs/1805.07278} {arXiv:1805.07278} \BibitemShut {NoStop}%
\bibitem [{\citenamefont {Skara}\ and\ \citenamefont
  {Perivolaropoulos}(2019)}]{leandros}%
  \BibitemOpen
  \bibfield  {author} {\bibinfo {author} {\bibfnamefont {F.}~\bibnamefont
  {Skara}}\ and\ \bibinfo {author} {\bibfnamefont {L.}~\bibnamefont
  {Perivolaropoulos}},\ }\href {\doibase 10.1103/PhysRevD.100.123527}
  {\bibfield  {journal} {\bibinfo  {journal} {Phys. Rev. D}\ }\textbf {\bibinfo
  {volume} {100}},\ \bibinfo {pages} {123527} (\bibinfo {year} {2019})},\
  \Eprint {http://arxiv.org/abs/1907.12594} {arXiv:1907.12594} \BibitemShut
  {NoStop}%
\bibitem [{\citenamefont {Majumder}(2011)}]{majumder}%
  \BibitemOpen
  \bibfield  {author} {\bibinfo {author} {\bibfnamefont {B.}~\bibnamefont
  {Majumder}},\ }\href {\doibase 10.1007/s10509-011-0815-6} {\bibfield
  {journal} {\bibinfo  {journal} {Astrophys. Space Sci.}\ }\textbf {\bibinfo
  {volume} {336}},\ \bibinfo {pages} {331} (\bibinfo {year} {2011})},\ \Eprint
  {http://arxiv.org/abs/1105.2425} {arXiv:1105.2425} \BibitemShut {NoStop}%
\bibitem [{\citenamefont {Faraoni}(2011)}]{faraoni}%
  \BibitemOpen
  \bibfield  {author} {\bibinfo {author} {\bibfnamefont {V.}~\bibnamefont
  {Faraoni}},\ }\href {\doibase 10.1103/PhysRevD.84.024003} {\bibfield
  {journal} {\bibinfo  {journal} {Phys. Rev. D}\ }\textbf {\bibinfo {volume}
  {84}},\ \bibinfo {pages} {024003} (\bibinfo {year} {2011})},\ \Eprint
  {http://arxiv.org/abs/1106.4427} {arXiv:1106.4427} \BibitemShut {NoStop}%
\bibitem [{\citenamefont {Gibbons}\ and\ \citenamefont
  {Hawking}(1977)}]{gibbhawk}%
  \BibitemOpen
  \bibfield  {author} {\bibinfo {author} {\bibfnamefont {G.}~\bibnamefont
  {Gibbons}}\ and\ \bibinfo {author} {\bibfnamefont {S.}~\bibnamefont
  {Hawking}},\ }\href {\doibase 10.1103/PhysRevD.15.2738} {\bibfield  {journal}
  {\bibinfo  {journal} {Phys. Rev. D}\ }\textbf {\bibinfo {volume} {15}},\
  \bibinfo {pages} {2738} (\bibinfo {year} {1977})}\BibitemShut {NoStop}%
\bibitem [{\citenamefont {Cai}\ \emph {et~al.}(2009)\citenamefont {Cai},
  \citenamefont {Cao},\ and\ \citenamefont {Hu}}]{caihawking}%
  \BibitemOpen
  \bibfield  {author} {\bibinfo {author} {\bibfnamefont {R.-G.}\ \bibnamefont
  {Cai}}, \bibinfo {author} {\bibfnamefont {L.-M.}\ \bibnamefont {Cao}}, \ and\
  \bibinfo {author} {\bibfnamefont {Y.-P.}\ \bibnamefont {Hu}},\ }\href
  {\doibase 10.1088/0264-9381/26/15/155018} {\bibfield  {journal} {\bibinfo
  {journal} {Class. Quant. Grav.}\ }\textbf {\bibinfo {volume} {26}},\ \bibinfo
  {pages} {155018} (\bibinfo {year} {2009})},\ \Eprint
  {http://arxiv.org/abs/0809.1554} {arXiv:0809.1554} \BibitemShut {NoStop}%
\bibitem [{\citenamefont {Park}(2008)}]{park}%
  \BibitemOpen
  \bibfield  {author} {\bibinfo {author} {\bibfnamefont {M.-i.}\ \bibnamefont
  {Park}},\ }\href {\doibase 10.1016/j.physletb.2007.11.090} {\bibfield
  {journal} {\bibinfo  {journal} {Phys. Lett.}\ }\textbf {\bibinfo {volume}
  {B659}},\ \bibinfo {pages} {698} (\bibinfo {year} {2008})},\ \Eprint
  {http://arxiv.org/abs/0709.2307} {arXiv:0709.2307} \BibitemShut {NoStop}%
\bibitem [{\citenamefont {Awad}\ and\ \citenamefont {Ali}(2014)}]{awadali}%
  \BibitemOpen
  \bibfield  {author} {\bibinfo {author} {\bibfnamefont {A.}~\bibnamefont
  {Awad}}\ and\ \bibinfo {author} {\bibfnamefont {A.~F.}\ \bibnamefont {Ali}},\
  }\href {\doibase 10.1007/JHEP06(2014)093} {\bibfield  {journal} {\bibinfo
  {journal} {JHEP}\ }\textbf {\bibinfo {volume} {06}},\ \bibinfo {pages} {093}
  (\bibinfo {year} {2014})},\ \Eprint {http://arxiv.org/abs/1404.7825}
  {arXiv:1404.7825} \BibitemShut {NoStop}%
\bibitem [{\citenamefont {Bull}\ \emph {et~al.}(2016)\citenamefont {Bull} \emph
  {et~al.}}]{Bull:2015stt}%
  \BibitemOpen
  \bibfield  {author} {\bibinfo {author} {\bibfnamefont {P.}~\bibnamefont
  {Bull}} \emph {et~al.},\ }\href {\doibase 10.1016/j.dark.2016.02.001}
  {\bibfield  {journal} {\bibinfo  {journal} {Phys. Dark Univ.}\ }\textbf
  {\bibinfo {volume} {12}},\ \bibinfo {pages} {56} (\bibinfo {year} {2016})},\
  \Eprint {http://arxiv.org/abs/1512.05356} {arXiv:1512.05356} \BibitemShut
  {NoStop}%
\bibitem [{\citenamefont {Vagenas}\ \emph {et~al.}(2018)\citenamefont
  {Vagenas}, \citenamefont {Alasfar}, \citenamefont {Alsaleh},\ and\
  \citenamefont {Ali}}]{Alasfar:2017loh}%
  \BibitemOpen
  \bibfield  {author} {\bibinfo {author} {\bibfnamefont {E.~C.}\ \bibnamefont
  {Vagenas}}, \bibinfo {author} {\bibfnamefont {L.}~\bibnamefont {Alasfar}},
  \bibinfo {author} {\bibfnamefont {S.~M.}\ \bibnamefont {Alsaleh}}, \ and\
  \bibinfo {author} {\bibfnamefont {A.~F.}\ \bibnamefont {Ali}},\ }\href
  {\doibase 10.1016/j.nuclphysb.2018.04.004} {\bibfield  {journal} {\bibinfo
  {journal} {Nucl. Phys. B}\ }\textbf {\bibinfo {volume} {931}},\ \bibinfo
  {pages} {72} (\bibinfo {year} {2018})},\ \Eprint
  {http://arxiv.org/abs/1706.06502} {arXiv:1706.06502} \BibitemShut {NoStop}%
\bibitem [{\citenamefont {Jizba}\ \emph {et~al.}(2010)\citenamefont {Jizba},
  \citenamefont {Kleinert},\ and\ \citenamefont {Scardigli}}]{Jizba:2009qf}%
  \BibitemOpen
  \bibfield  {author} {\bibinfo {author} {\bibfnamefont {P.}~\bibnamefont
  {Jizba}}, \bibinfo {author} {\bibfnamefont {H.}~\bibnamefont {Kleinert}}, \
  and\ \bibinfo {author} {\bibfnamefont {F.}~\bibnamefont {Scardigli}},\ }\href
  {\doibase 10.1103/PhysRevD.81.084030} {\bibfield  {journal} {\bibinfo
  {journal} {Phys. Rev. D}\ }\textbf {\bibinfo {volume} {81}},\ \bibinfo
  {pages} {084030} (\bibinfo {year} {2010})},\ \Eprint
  {http://arxiv.org/abs/0912.2253} {arXiv:0912.2253} \BibitemShut {NoStop}%
\bibitem [{\citenamefont {Ong}\ and\ \citenamefont {Yao}(2018)}]{Ong:2018nzk}%
  \BibitemOpen
  \bibfield  {author} {\bibinfo {author} {\bibfnamefont {Y.~C.}\ \bibnamefont
  {Ong}}\ and\ \bibinfo {author} {\bibfnamefont {Y.}~\bibnamefont {Yao}},\
  }\href {\doibase 10.1103/PhysRevD.98.126018} {\bibfield  {journal} {\bibinfo
  {journal} {Phys. Rev. D}\ }\textbf {\bibinfo {volume} {98}},\ \bibinfo
  {pages} {126018} (\bibinfo {year} {2018})},\ \Eprint
  {http://arxiv.org/abs/1809.06348} {arXiv:1809.06348} \BibitemShut {NoStop}%
\bibitem [{\citenamefont {Buoninfante}\ \emph {et~al.}(2019)\citenamefont
  {Buoninfante}, \citenamefont {Luciano},\ and\ \citenamefont
  {Petruzziello}}]{Buoninfante:2019fwr}%
  \BibitemOpen
  \bibfield  {author} {\bibinfo {author} {\bibfnamefont {L.}~\bibnamefont
  {Buoninfante}}, \bibinfo {author} {\bibfnamefont {G.~G.}\ \bibnamefont
  {Luciano}}, \ and\ \bibinfo {author} {\bibfnamefont {L.}~\bibnamefont
  {Petruzziello}},\ }\href {\doibase 10.1140/epjc/s10052-019-7164-y} {\bibfield
   {journal} {\bibinfo  {journal} {Eur. Phys. J. C}\ }\textbf {\bibinfo
  {volume} {79}},\ \bibinfo {pages} {663} (\bibinfo {year} {2019})},\ \Eprint
  {http://arxiv.org/abs/1903.01382} {arXiv:1903.01382} \BibitemShut {NoStop}%
\bibitem [{\citenamefont {Ong}(2018)}]{Ong:2018zqn}%
  \BibitemOpen
  \bibfield  {author} {\bibinfo {author} {\bibfnamefont {Y.~C.}\ \bibnamefont
  {Ong}},\ }\href {\doibase 10.1088/1475-7516/2018/09/015} {\bibfield
  {journal} {\bibinfo  {journal} {JCAP}\ }\textbf {\bibinfo {volume} {09}},\
  \bibinfo {pages} {015} (\bibinfo {year} {2018})},\ \Eprint
  {http://arxiv.org/abs/1804.05176} {arXiv:1804.05176} \BibitemShut {NoStop}%
\bibitem [{\citenamefont {Tawfik}\ and\ \citenamefont {Diab}(2014)}]{tawfik}%
  \BibitemOpen
  \bibfield  {author} {\bibinfo {author} {\bibfnamefont {A.~N.}\ \bibnamefont
  {Tawfik}}\ and\ \bibinfo {author} {\bibfnamefont {A.~M.}\ \bibnamefont
  {Diab}},\ }\href {\doibase 10.1142/S0218271814300250} {\bibfield  {journal}
  {\bibinfo  {journal} {Int. J. Mod. Phys. D}\ }\textbf {\bibinfo {volume}
  {23}},\ \bibinfo {pages} {1430025} (\bibinfo {year} {2014})},\ \Eprint
  {http://arxiv.org/abs/1410.0206} {arXiv:1410.0206} \BibitemShut {NoStop}%
\bibitem [{\citenamefont {Tawfik}\ and\ \citenamefont {Diab}(2015)}]{diab}%
  \BibitemOpen
  \bibfield  {author} {\bibinfo {author} {\bibfnamefont {A.~N.}\ \bibnamefont
  {Tawfik}}\ and\ \bibinfo {author} {\bibfnamefont {A.~M.}\ \bibnamefont
  {Diab}},\ }\href {\doibase 10.1088/0034-4885/78/12/126001} {\bibfield
  {journal} {\bibinfo  {journal} {Rept. Prog. Phys.}\ }\textbf {\bibinfo
  {volume} {78}},\ \bibinfo {pages} {126001} (\bibinfo {year} {2015})},\
  \Eprint {http://arxiv.org/abs/1509.02436} {arXiv:1509.02436} \BibitemShut
  {NoStop}%
\bibitem [{\citenamefont {Scardigli}(2019)}]{scardrev}%
  \BibitemOpen
  \bibfield  {author} {\bibinfo {author} {\bibfnamefont {F.}~\bibnamefont
  {Scardigli}},\ }\href {\doibase 10.1088/1742-6596/1275/1/012004} {\bibfield
  {journal} {\bibinfo  {journal} {J. Phys. Conf. Ser.}\ }\textbf {\bibinfo
  {volume} {1275}},\ \bibinfo {pages} {012004} (\bibinfo {year} {2019})},\
  \Eprint {http://arxiv.org/abs/1905.00287} {arXiv:1905.00287} \BibitemShut
  {NoStop}%
\bibitem [{\citenamefont {Scardigli}\ and\ \citenamefont
  {Casadio}(2015)}]{casadio}%
  \BibitemOpen
  \bibfield  {author} {\bibinfo {author} {\bibfnamefont {F.}~\bibnamefont
  {Scardigli}}\ and\ \bibinfo {author} {\bibfnamefont {R.}~\bibnamefont
  {Casadio}},\ }\href {\doibase 10.1140/epjc/s10052-015-3635-y} {\bibfield
  {journal} {\bibinfo  {journal} {Eur. Phys. J. C}\ }\textbf {\bibinfo {volume}
  {75}},\ \bibinfo {pages} {425} (\bibinfo {year} {2015})},\ \Eprint
  {http://arxiv.org/abs/1407.0113} {arXiv:1407.0113} \BibitemShut {NoStop}%
\bibitem [{\citenamefont {Feng}\ \emph {et~al.}(2017)\citenamefont {Feng},
  \citenamefont {Yang}, \citenamefont {Li},\ and\ \citenamefont {Zu}}]{gw}%
  \BibitemOpen
  \bibfield  {author} {\bibinfo {author} {\bibfnamefont {Z.-W.}\ \bibnamefont
  {Feng}}, \bibinfo {author} {\bibfnamefont {S.-Z.}\ \bibnamefont {Yang}},
  \bibinfo {author} {\bibfnamefont {H.-L.}\ \bibnamefont {Li}}, \ and\ \bibinfo
  {author} {\bibfnamefont {X.-T.}\ \bibnamefont {Zu}},\ }\href {\doibase
  10.1016/j.physletb.2017.02.043} {\bibfield  {journal} {\bibinfo  {journal}
  {Phys. Lett. B}\ }\textbf {\bibinfo {volume} {768}},\ \bibinfo {pages} {81}
  (\bibinfo {year} {2017})},\ \Eprint {http://arxiv.org/abs/1610.08549}
  {arXiv:1610.08549} \BibitemShut {NoStop}%
\bibitem [{\citenamefont {Ghosh}(2014)}]{ghosh}%
  \BibitemOpen
  \bibfield  {author} {\bibinfo {author} {\bibfnamefont {S.}~\bibnamefont
  {Ghosh}},\ }\href {\doibase 10.1088/0264-9381/31/2/025025} {\bibfield
  {journal} {\bibinfo  {journal} {Class. Quant. Grav.}\ }\textbf {\bibinfo
  {volume} {31}},\ \bibinfo {pages} {025025} (\bibinfo {year} {2014})},\
  \Eprint {http://arxiv.org/abs/1303.1256} {arXiv:1303.1256} \BibitemShut
  {NoStop}%
\bibitem [{\citenamefont {Gao}\ \emph {et~al.}(2017)\citenamefont {Gao},
  \citenamefont {Wang},\ and\ \citenamefont {Zhan}}]{Gao:2017zch}%
  \BibitemOpen
  \bibfield  {author} {\bibinfo {author} {\bibfnamefont {D.}~\bibnamefont
  {Gao}}, \bibinfo {author} {\bibfnamefont {J.}~\bibnamefont {Wang}}, \ and\
  \bibinfo {author} {\bibfnamefont {M.}~\bibnamefont {Zhan}},\ }\href {\doibase
  10.1103/PhysRevA.95.042106} {\bibfield  {journal} {\bibinfo  {journal} {Phys.
  Rev. A}\ }\textbf {\bibinfo {volume} {95}},\ \bibinfo {pages} {042106}
  (\bibinfo {year} {2017})},\ \Eprint {http://arxiv.org/abs/1704.02037}
  {arXiv:1704.02037} \BibitemShut {NoStop}%
\bibitem [{\citenamefont {Ali}\ \emph {et~al.}(2011)\citenamefont {Ali},
  \citenamefont {Das},\ and\ \citenamefont {Vagenas}}]{qglab}%
  \BibitemOpen
  \bibfield  {author} {\bibinfo {author} {\bibfnamefont {A.~F.}\ \bibnamefont
  {Ali}}, \bibinfo {author} {\bibfnamefont {S.}~\bibnamefont {Das}}, \ and\
  \bibinfo {author} {\bibfnamefont {E.~C.}\ \bibnamefont {Vagenas}},\ }\href
  {\doibase 10.1103/PhysRevD.84.044013} {\bibfield  {journal} {\bibinfo
  {journal} {Phys. Rev. D}\ }\textbf {\bibinfo {volume} {84}},\ \bibinfo
  {pages} {044013} (\bibinfo {year} {2011})},\ \Eprint
  {http://arxiv.org/abs/1107.3164} {arXiv:1107.3164} \BibitemShut {NoStop}%
\bibitem [{\citenamefont {Das}\ and\ \citenamefont
  {Vagenas}(2008)}]{Das:2008kaa}%
  \BibitemOpen
  \bibfield  {author} {\bibinfo {author} {\bibfnamefont {S.}~\bibnamefont
  {Das}}\ and\ \bibinfo {author} {\bibfnamefont {E.~C.}\ \bibnamefont
  {Vagenas}},\ }\href {\doibase 10.1103/PhysRevLett.101.221301} {\bibfield
  {journal} {\bibinfo  {journal} {Phys. Rev. Lett.}\ }\textbf {\bibinfo
  {volume} {101}},\ \bibinfo {pages} {221301} (\bibinfo {year} {2008})},\
  \Eprint {http://arxiv.org/abs/0810.5333} {arXiv:0810.5333} \BibitemShut
  {NoStop}%
\bibitem [{\citenamefont {Das}\ and\ \citenamefont
  {Vagenas}(2009)}]{Das:2009hs}%
  \BibitemOpen
  \bibfield  {author} {\bibinfo {author} {\bibfnamefont {S.}~\bibnamefont
  {Das}}\ and\ \bibinfo {author} {\bibfnamefont {E.~C.}\ \bibnamefont
  {Vagenas}},\ }\href {\doibase 10.1139/P08-105} {\bibfield  {journal}
  {\bibinfo  {journal} {Can. J. Phys.}\ }\textbf {\bibinfo {volume} {87}},\
  \bibinfo {pages} {233} (\bibinfo {year} {2009})},\ \Eprint
  {http://arxiv.org/abs/0901.1768} {arXiv:0901.1768} \BibitemShut {NoStop}%
\bibitem [{\citenamefont {Bushev}\ \emph {et~al.}(2019)\citenamefont {Bushev},
  \citenamefont {Bourhill}, \citenamefont {Goryachev}, \citenamefont
  {Kukharchyk}, \citenamefont {Ivanov}, \citenamefont {Galliou}, \citenamefont
  {Tobar},\ and\ \citenamefont {Danilishin}}]{oscillators}%
  \BibitemOpen
  \bibfield  {author} {\bibinfo {author} {\bibfnamefont {P.}~\bibnamefont
  {Bushev}}, \bibinfo {author} {\bibfnamefont {J.}~\bibnamefont {Bourhill}},
  \bibinfo {author} {\bibfnamefont {M.}~\bibnamefont {Goryachev}}, \bibinfo
  {author} {\bibfnamefont {N.}~\bibnamefont {Kukharchyk}}, \bibinfo {author}
  {\bibfnamefont {E.}~\bibnamefont {Ivanov}}, \bibinfo {author} {\bibfnamefont
  {S.}~\bibnamefont {Galliou}}, \bibinfo {author} {\bibfnamefont
  {M.}~\bibnamefont {Tobar}}, \ and\ \bibinfo {author} {\bibfnamefont
  {S.}~\bibnamefont {Danilishin}},\ }\href {\doibase
  10.1103/PhysRevD.100.066020} {\bibfield  {journal} {\bibinfo  {journal}
  {Phys. Rev. D}\ }\textbf {\bibinfo {volume} {100}},\ \bibinfo {pages}
  {066020} (\bibinfo {year} {2019})},\ \Eprint
  {http://arxiv.org/abs/1903.03346} {arXiv:1903.03346} \BibitemShut {NoStop}%
\bibitem [{\citenamefont {Das}\ and\ \citenamefont {Mann}(2011)}]{Das:2011tq}%
  \BibitemOpen
  \bibfield  {author} {\bibinfo {author} {\bibfnamefont {S.}~\bibnamefont
  {Das}}\ and\ \bibinfo {author} {\bibfnamefont {R.}~\bibnamefont {Mann}},\
  }\href {\doibase 10.1016/j.physletb.2011.09.056} {\bibfield  {journal}
  {\bibinfo  {journal} {Phys. Lett. B}\ }\textbf {\bibinfo {volume} {704}},\
  \bibinfo {pages} {596} (\bibinfo {year} {2011})},\ \Eprint
  {http://arxiv.org/abs/1109.3258} {arXiv:1109.3258} \BibitemShut {NoStop}%
\bibitem [{\citenamefont {Marin}\ \emph {et~al.}(2013)\citenamefont {Marin}
  \emph {et~al.}}]{Marin:2013pga}%
  \BibitemOpen
  \bibfield  {author} {\bibinfo {author} {\bibfnamefont {F.}~\bibnamefont
  {Marin}} \emph {et~al.},\ }\href {\doibase 10.1038/nphys2503} {\bibfield
  {journal} {\bibinfo  {journal} {Nature Phys.}\ }\textbf {\bibinfo {volume}
  {9}},\ \bibinfo {pages} {71} (\bibinfo {year} {2013})}\BibitemShut {NoStop}%
\bibitem [{\citenamefont {Gao}\ and\ \citenamefont {Zhan}(2016)}]{Gao:2016fmk}%
  \BibitemOpen
  \bibfield  {author} {\bibinfo {author} {\bibfnamefont {D.}~\bibnamefont
  {Gao}}\ and\ \bibinfo {author} {\bibfnamefont {M.}~\bibnamefont {Zhan}},\
  }\href {\doibase 10.1103/PhysRevA.94.013607} {\bibfield  {journal} {\bibinfo
  {journal} {Phys. Rev. A}\ }\textbf {\bibinfo {volume} {94}},\ \bibinfo
  {pages} {013607} (\bibinfo {year} {2016})},\ \Eprint
  {http://arxiv.org/abs/1607.04353} {arXiv:1607.04353} \BibitemShut {NoStop}%
\bibitem [{\citenamefont {Neves}(2020)}]{neves}%
  \BibitemOpen
  \bibfield  {author} {\bibinfo {author} {\bibfnamefont {J.~C.}\ \bibnamefont
  {Neves}},\ }\href {\doibase 10.1140/epjc/s10052-020-7913-y} {\bibfield
  {journal} {\bibinfo  {journal} {Eur. Phys. J. C}\ }\textbf {\bibinfo {volume}
  {80}},\ \bibinfo {pages} {343} (\bibinfo {year} {2020})},\ \Eprint
  {http://arxiv.org/abs/1906.11735} {arXiv:1906.11735} \BibitemShut {NoStop}%
\bibitem [{\citenamefont {Wong}\ \emph {et~al.}(2019)\citenamefont {Wong} \emph
  {et~al.}}]{Wong:2019kwg}%
  \BibitemOpen
  \bibfield  {author} {\bibinfo {author} {\bibfnamefont {K.~C.}\ \bibnamefont
  {Wong}} \emph {et~al.},\ }\href@noop {} {\  (\bibinfo {year} {2019})},\
  \Eprint {http://arxiv.org/abs/1907.04869} {arXiv:1907.04869} \BibitemShut
  {NoStop}%
\bibitem [{\citenamefont {Berg}(2004)}]{Berg}%
  \BibitemOpen
  \bibfield  {author} {\bibinfo {author} {\bibfnamefont {B.~A.}\ \bibnamefont
  {Berg}},\ }\href@noop {} {\emph {\bibinfo {title} {{Markov Chain Monte Carlo
  Simulations and Their Statistical Analysis}}}}\ (\bibinfo  {publisher} {World
  Scientific, Singapore},\ \bibinfo {year} {2004})\BibitemShut {NoStop}%
\bibitem [{\citenamefont {MacKay}(2003)}]{MacKay}%
  \BibitemOpen
  \bibfield  {author} {\bibinfo {author} {\bibfnamefont {D.~J.~C.}\
  \bibnamefont {MacKay}},\ }\href@noop {} {\emph {\bibinfo {title}
  {{Information Theory, Inference, and Learning Algorithms}}}}\ (\bibinfo
  {publisher} {Cambridge University Press},\ \bibinfo {year}
  {2003})\BibitemShut {NoStop}%
\bibitem [{\citenamefont {Neal}(1993)}]{Neal}%
  \BibitemOpen
  \bibfield  {author} {\bibinfo {author} {\bibfnamefont {R.~M.}\ \bibnamefont
  {Neal}},\ }\href@noop {} {\emph {\bibinfo {title} {{Probabilistic Inference
  Using Markoc Chain Monte Carlo Methods}}}}\ (\bibinfo  {publisher}
  {Department of Computer Science, University of Toronto},\ \bibinfo {year}
  {1993})\BibitemShut {NoStop}%
\bibitem [{\citenamefont {Dunkley}\ \emph {et~al.}(2005)\citenamefont
  {Dunkley}, \citenamefont {Bucher}, \citenamefont {Ferreira}, \citenamefont
  {Moodley},\ and\ \citenamefont {Skordis}}]{Dunkley:2004sv}%
  \BibitemOpen
  \bibfield  {author} {\bibinfo {author} {\bibfnamefont {J.}~\bibnamefont
  {Dunkley}}, \bibinfo {author} {\bibfnamefont {M.}~\bibnamefont {Bucher}},
  \bibinfo {author} {\bibfnamefont {P.~G.}\ \bibnamefont {Ferreira}}, \bibinfo
  {author} {\bibfnamefont {K.}~\bibnamefont {Moodley}}, \ and\ \bibinfo
  {author} {\bibfnamefont {C.}~\bibnamefont {Skordis}},\ }\href {\doibase
  10.1111/j.1365-2966.2004.08464.x} {\bibfield  {journal} {\bibinfo  {journal}
  {Mon. Not. Roy. Astron. Soc.}\ }\textbf {\bibinfo {volume} {356}},\ \bibinfo
  {pages} {925} (\bibinfo {year} {2005})},\ \Eprint
  {http://arxiv.org/abs/astro-ph/0405462} {arXiv:astro-ph/0405462} \BibitemShut
  {NoStop}%
\bibitem [{\citenamefont {Mukherjee}\ \emph {et~al.}(2006)\citenamefont
  {Mukherjee}, \citenamefont {Parkinson},\ and\ \citenamefont
  {Liddle}}]{Mukherjee:2005wg}%
  \BibitemOpen
  \bibfield  {author} {\bibinfo {author} {\bibfnamefont {P.}~\bibnamefont
  {Mukherjee}}, \bibinfo {author} {\bibfnamefont {D.}~\bibnamefont
  {Parkinson}}, \ and\ \bibinfo {author} {\bibfnamefont {A.~R.}\ \bibnamefont
  {Liddle}},\ }\href {\doibase 10.1086/501068} {\bibfield  {journal} {\bibinfo
  {journal} {Astrophys. J. Lett.}\ }\textbf {\bibinfo {volume} {638}},\
  \bibinfo {pages} {L51} (\bibinfo {year} {2006})},\ \Eprint
  {http://arxiv.org/abs/astro-ph/0508461} {arXiv:astro-ph/0508461} \BibitemShut
  {NoStop}%
\bibitem [{\citenamefont {Nesseris}\ and\ \citenamefont
  {Garcia-Bellido}(2013)}]{Nesseris:2012cq}%
  \BibitemOpen
  \bibfield  {author} {\bibinfo {author} {\bibfnamefont {S.}~\bibnamefont
  {Nesseris}}\ and\ \bibinfo {author} {\bibfnamefont {J.}~\bibnamefont
  {Garcia-Bellido}},\ }\href {\doibase 10.1088/1475-7516/2013/08/036}
  {\bibfield  {journal} {\bibinfo  {journal} {JCAP}\ }\textbf {\bibinfo
  {volume} {08}},\ \bibinfo {pages} {036} (\bibinfo {year} {2013})},\ \Eprint
  {http://arxiv.org/abs/1210.7652} {arXiv:1210.7652} \BibitemShut {NoStop}%
\bibitem [{\citenamefont {Jeffreys}(1939)}]{Jeffreys:1939xee}%
  \BibitemOpen
  \bibfield  {author} {\bibinfo {author} {\bibfnamefont {H.}~\bibnamefont
  {Jeffreys}},\ }\href@noop {} {\emph {\bibinfo {title} {{The Theory of
  Probability}}}}\ (\bibinfo  {publisher} {Oxford Classic Texts in the Physical
  Sciences},\ \bibinfo {year} {1939})\BibitemShut {NoStop}%
\bibitem [{\citenamefont {Scolnic}\ \emph {et~al.}(2018)\citenamefont {Scolnic}
  \emph {et~al.}}]{Scolnic:2017caz}%
  \BibitemOpen
  \bibfield  {author} {\bibinfo {author} {\bibfnamefont {D.}~\bibnamefont
  {Scolnic}} \emph {et~al.},\ }\href {\doibase 10.3847/1538-4357/aab9bb}
  {\bibfield  {journal} {\bibinfo  {journal} {Astrophys. J.}\ }\textbf
  {\bibinfo {volume} {859}},\ \bibinfo {pages} {101} (\bibinfo {year}
  {2018})},\ \Eprint {http://arxiv.org/abs/1710.00845} {arXiv:1710.00845}
  \BibitemShut {NoStop}%
\bibitem [{\citenamefont {Conley}\ \emph {et~al.}(2011)\citenamefont {Conley}
  \emph {et~al.}}]{conley}%
  \BibitemOpen
  \bibfield  {author} {\bibinfo {author} {\bibfnamefont {A.}~\bibnamefont
  {Conley}} \emph {et~al.} (\bibinfo {collaboration} {SNLS}),\ }\href {\doibase
  10.1088/0067-0049/192/1/1} {\bibfield  {journal} {\bibinfo  {journal}
  {Astrophys. J. Suppl.}\ }\textbf {\bibinfo {volume} {192}},\ \bibinfo {pages}
  {1} (\bibinfo {year} {2011})},\ \Eprint {http://arxiv.org/abs/1104.1443}
  {arXiv:1104.1443} \BibitemShut {NoStop}%
\bibitem [{\citenamefont {Moresco}(2015)}]{moresco}%
  \BibitemOpen
  \bibfield  {author} {\bibinfo {author} {\bibfnamefont {M.}~\bibnamefont
  {Moresco}},\ }\href {\doibase 10.1093/mnrasl/slv037} {\bibfield  {journal}
  {\bibinfo  {journal} {Mon. Not. Roy. Astron. Soc.}\ }\textbf {\bibinfo
  {volume} {450}},\ \bibinfo {pages} {L16} (\bibinfo {year} {2015})},\ \Eprint
  {http://arxiv.org/abs/1503.01116} {arXiv:1503.01116} \BibitemShut {NoStop}%
\bibitem [{\citenamefont {Suyu}\ \emph {et~al.}(2017)\citenamefont {Suyu} \emph
  {et~al.}}]{Suyu:2016qxx}%
  \BibitemOpen
  \bibfield  {author} {\bibinfo {author} {\bibfnamefont {S.}~\bibnamefont
  {Suyu}} \emph {et~al.},\ }\href {\doibase 10.1093/mnras/stx483} {\bibfield
  {journal} {\bibinfo  {journal} {Mon. Not. Roy. Astron. Soc.}\ }\textbf
  {\bibinfo {volume} {468}},\ \bibinfo {pages} {2590} (\bibinfo {year}
  {2017})},\ \Eprint {http://arxiv.org/abs/1607.00017} {arXiv:1607.00017}
  \BibitemShut {NoStop}%
\bibitem [{\citenamefont {{Schneider}}\ \emph {et~al.}(1992)\citenamefont
  {{Schneider}}, \citenamefont {{Ehlers}},\ and\ \citenamefont
  {{Falco}}}]{gralen.boo}%
  \BibitemOpen
  \bibfield  {author} {\bibinfo {author} {\bibfnamefont {P.}~\bibnamefont
  {{Schneider}}}, \bibinfo {author} {\bibfnamefont {J.}~\bibnamefont
  {{Ehlers}}}, \ and\ \bibinfo {author} {\bibfnamefont {E.~E.}\ \bibnamefont
  {{Falco}}},\ }\href@noop {} {\emph {\bibinfo {title} {{Gravitational
  Lenses}}}}\ (\bibinfo  {publisher} {Springer-Verlag Berlin Heidelberg New
  York},\ \bibinfo {year} {1992})\BibitemShut {NoStop}%
\bibitem [{\citenamefont {Hogg}(1999)}]{Hogg:1999ad}%
  \BibitemOpen
  \bibfield  {author} {\bibinfo {author} {\bibfnamefont {D.~W.}\ \bibnamefont
  {Hogg}},\ }\href@noop {} {\  (\bibinfo {year} {1999})},\ \Eprint
  {http://arxiv.org/abs/astro-ph/9905116} {arXiv:astro-ph/9905116} \BibitemShut
  {NoStop}%
\bibitem [{\citenamefont {Liu}\ and\ \citenamefont {Wei}(2015)}]{Liu:2014vda}%
  \BibitemOpen
  \bibfield  {author} {\bibinfo {author} {\bibfnamefont {J.}~\bibnamefont
  {Liu}}\ and\ \bibinfo {author} {\bibfnamefont {H.}~\bibnamefont {Wei}},\
  }\href {\doibase 10.1007/s10714-015-1986-1} {\bibfield  {journal} {\bibinfo
  {journal} {Gen. Rel. Grav.}\ }\textbf {\bibinfo {volume} {47}},\ \bibinfo
  {pages} {141} (\bibinfo {year} {2015})},\ \Eprint
  {http://arxiv.org/abs/1410.3960} {arXiv:1410.3960} \BibitemShut {NoStop}%
\bibitem [{\citenamefont {Blake}\ \emph {et~al.}(2012)\citenamefont {Blake}
  \emph {et~al.}}]{Blake:2012pj}%
  \BibitemOpen
  \bibfield  {author} {\bibinfo {author} {\bibfnamefont {C.}~\bibnamefont
  {Blake}} \emph {et~al.},\ }\href {\doibase 10.1111/j.1365-2966.2012.21473.x}
  {\bibfield  {journal} {\bibinfo  {journal} {Mon. Not. Roy. Astron. Soc.}\
  }\textbf {\bibinfo {volume} {425}},\ \bibinfo {pages} {405} (\bibinfo {year}
  {2012})},\ \Eprint {http://arxiv.org/abs/1204.3674} {arXiv:1204.3674}
  \BibitemShut {NoStop}%
\bibitem [{\citenamefont {Alam}\ \emph {et~al.}(2017)\citenamefont {Alam} \emph
  {et~al.}}]{Alam:2016hwk}%
  \BibitemOpen
  \bibfield  {author} {\bibinfo {author} {\bibfnamefont {S.}~\bibnamefont
  {Alam}} \emph {et~al.} (\bibinfo {collaboration} {BOSS}),\ }\href {\doibase
  10.1093/mnras/stx721} {\bibfield  {journal} {\bibinfo  {journal} {Mon. Not.
  Roy. Astron. Soc.}\ }\textbf {\bibinfo {volume} {470}},\ \bibinfo {pages}
  {2617} (\bibinfo {year} {2017})},\ \Eprint {http://arxiv.org/abs/1607.03155}
  {arXiv:1607.03155} \BibitemShut {NoStop}%
\bibitem [{\citenamefont {Eisenstein}\ and\ \citenamefont
  {Hu}(1998)}]{Eisenstein:1997ik}%
  \BibitemOpen
  \bibfield  {author} {\bibinfo {author} {\bibfnamefont {D.~J.}\ \bibnamefont
  {Eisenstein}}\ and\ \bibinfo {author} {\bibfnamefont {W.}~\bibnamefont
  {Hu}},\ }\href {\doibase 10.1086/305424} {\bibfield  {journal} {\bibinfo
  {journal} {Astrophys. J.}\ }\textbf {\bibinfo {volume} {496}},\ \bibinfo
  {pages} {605} (\bibinfo {year} {1998})},\ \Eprint
  {http://arxiv.org/abs/astro-ph/9709112} {arXiv:astro-ph/9709112} \BibitemShut
  {NoStop}%
\bibitem [{\citenamefont {Nadathur}\ \emph {et~al.}(2019)\citenamefont
  {Nadathur}, \citenamefont {Carter}, \citenamefont {Percival}, \citenamefont
  {Winther},\ and\ \citenamefont {Bautista}}]{Nadathur:2019mct}%
  \BibitemOpen
  \bibfield  {author} {\bibinfo {author} {\bibfnamefont {S.}~\bibnamefont
  {Nadathur}}, \bibinfo {author} {\bibfnamefont {P.~M.}\ \bibnamefont
  {Carter}}, \bibinfo {author} {\bibfnamefont {W.~J.}\ \bibnamefont
  {Percival}}, \bibinfo {author} {\bibfnamefont {H.~A.}\ \bibnamefont
  {Winther}}, \ and\ \bibinfo {author} {\bibfnamefont {J.}~\bibnamefont
  {Bautista}},\ }\href {\doibase 10.1103/PhysRevD.100.023504} {\bibfield
  {journal} {\bibinfo  {journal} {Phys. Rev. D}\ }\textbf {\bibinfo {volume}
  {100}},\ \bibinfo {pages} {023504} (\bibinfo {year} {2019})},\ \Eprint
  {http://arxiv.org/abs/1904.01030} {arXiv:1904.01030} \BibitemShut {NoStop}%
\bibitem [{\citenamefont {Ata}\ \emph {et~al.}(2018)\citenamefont {Ata} \emph
  {et~al.}}]{Ata:2017dya}%
  \BibitemOpen
  \bibfield  {author} {\bibinfo {author} {\bibfnamefont {M.}~\bibnamefont
  {Ata}} \emph {et~al.},\ }\href {\doibase 10.1093/mnras/stx2630} {\bibfield
  {journal} {\bibinfo  {journal} {Mon. Not. Roy. Astron. Soc.}\ }\textbf
  {\bibinfo {volume} {473}},\ \bibinfo {pages} {4773} (\bibinfo {year}
  {2018})},\ \Eprint {http://arxiv.org/abs/1705.06373} {arXiv:1705.06373}
  \BibitemShut {NoStop}%
\bibitem [{\citenamefont {de~Sainte~Agathe}\ \emph {et~al.}(2019)\citenamefont
  {de~Sainte~Agathe} \emph {et~al.}}]{lyman}%
  \BibitemOpen
  \bibfield  {author} {\bibinfo {author} {\bibfnamefont {V.}~\bibnamefont
  {de~Sainte~Agathe}} \emph {et~al.},\ }\href {\doibase
  10.1051/0004-6361/201935638} {\bibfield  {journal} {\bibinfo  {journal}
  {Astron. Astrophys.}\ }\textbf {\bibinfo {volume} {629}},\ \bibinfo {pages}
  {A85} (\bibinfo {year} {2019})},\ \Eprint {http://arxiv.org/abs/1904.03400}
  {arXiv:1904.03400} \BibitemShut {NoStop}%
\bibitem [{\citenamefont {Blomqvist}\ \emph {et~al.}(2019)\citenamefont
  {Blomqvist} \emph {et~al.}}]{Blomqvist:2019rah}%
  \BibitemOpen
  \bibfield  {author} {\bibinfo {author} {\bibfnamefont {M.}~\bibnamefont
  {Blomqvist}} \emph {et~al.},\ }\href {\doibase 10.1051/0004-6361/201935641}
  {\bibfield  {journal} {\bibinfo  {journal} {Astron. Astrophys.}\ }\textbf
  {\bibinfo {volume} {629}},\ \bibinfo {pages} {A86} (\bibinfo {year}
  {2019})},\ \Eprint {http://arxiv.org/abs/1904.03430} {arXiv:1904.03430}
  \BibitemShut {NoStop}%
\bibitem [{\citenamefont {Wang}\ and\ \citenamefont
  {Mukherjee}(2007)}]{Wang:2007mza}%
  \BibitemOpen
  \bibfield  {author} {\bibinfo {author} {\bibfnamefont {Y.}~\bibnamefont
  {Wang}}\ and\ \bibinfo {author} {\bibfnamefont {P.}~\bibnamefont
  {Mukherjee}},\ }\href {\doibase 10.1103/PhysRevD.76.103533} {\bibfield
  {journal} {\bibinfo  {journal} {Phys. Rev. D}\ }\textbf {\bibinfo {volume}
  {76}},\ \bibinfo {pages} {103533} (\bibinfo {year} {2007})},\ \Eprint
  {http://arxiv.org/abs/astro-ph/0703780} {arXiv:astro-ph/0703780} \BibitemShut
  {NoStop}%
\bibitem [{\citenamefont {Zhai}\ and\ \citenamefont {Wang}(2019)}]{cmb&sn}%
  \BibitemOpen
  \bibfield  {author} {\bibinfo {author} {\bibfnamefont {Z.}~\bibnamefont
  {Zhai}}\ and\ \bibinfo {author} {\bibfnamefont {Y.}~\bibnamefont {Wang}},\
  }\href {\doibase 10.1088/1475-7516/2019/07/005} {\bibfield  {journal}
  {\bibinfo  {journal} {JCAP}\ }\textbf {\bibinfo {volume} {07}},\ \bibinfo
  {pages} {005} (\bibinfo {year} {2019})},\ \Eprint
  {http://arxiv.org/abs/1811.07425} {arXiv:1811.07425} \BibitemShut {NoStop}%
\bibitem [{\citenamefont {Hu}\ and\ \citenamefont
  {Sugiyama}(1996)}]{Hu:1995en}%
  \BibitemOpen
  \bibfield  {author} {\bibinfo {author} {\bibfnamefont {W.}~\bibnamefont
  {Hu}}\ and\ \bibinfo {author} {\bibfnamefont {N.}~\bibnamefont {Sugiyama}},\
  }\href {\doibase 10.1086/177989} {\bibfield  {journal} {\bibinfo  {journal}
  {Astrophys. J.}\ }\textbf {\bibinfo {volume} {471}},\ \bibinfo {pages} {542}
  (\bibinfo {year} {1996})},\ \Eprint {http://arxiv.org/abs/astro-ph/9510117}
  {arXiv:astro-ph/9510117} \BibitemShut {NoStop}%
\bibitem [{\citenamefont {Aghanim}\ \emph {et~al.}(2018)\citenamefont {Aghanim}
  \emph {et~al.}}]{Aghanim:2018eyx}%
  \BibitemOpen
  \bibfield  {author} {\bibinfo {author} {\bibfnamefont {N.}~\bibnamefont
  {Aghanim}} \emph {et~al.} (\bibinfo {collaboration} {Planck}),\ }\href@noop
  {} {\  (\bibinfo {year} {2018})},\ \Eprint {http://arxiv.org/abs/1807.06209}
  {arXiv:1807.06209} \BibitemShut {NoStop}%
\bibitem [{\citenamefont {Di~Valentino}\ \emph {et~al.}(2019)\citenamefont
  {Di~Valentino}, \citenamefont {Melchiorri},\ and\ \citenamefont
  {Silk}}]{DiValentino:2019qzk}%
  \BibitemOpen
  \bibfield  {author} {\bibinfo {author} {\bibfnamefont {E.}~\bibnamefont
  {Di~Valentino}}, \bibinfo {author} {\bibfnamefont {A.}~\bibnamefont
  {Melchiorri}}, \ and\ \bibinfo {author} {\bibfnamefont {J.}~\bibnamefont
  {Silk}},\ }\href {\doibase 10.1038/s41550-019-0906-9} {\bibfield  {journal}
  {\bibinfo  {journal} {Nature Astron.}\ }\textbf {\bibinfo {volume} {4}},\
  \bibinfo {pages} {196} (\bibinfo {year} {2019})},\ \Eprint
  {http://arxiv.org/abs/1911.02087} {arXiv:1911.02087} \BibitemShut {NoStop}%
\bibitem [{\citenamefont {Handley}(2019)}]{handley}%
  \BibitemOpen
  \bibfield  {author} {\bibinfo {author} {\bibfnamefont {W.}~\bibnamefont
  {Handley}},\ }\href@noop {} {\  (\bibinfo {year} {2019})},\ \Eprint
  {http://arxiv.org/abs/1908.09139} {arXiv:1908.09139} \BibitemShut {NoStop}%
\bibitem [{\citenamefont {Maggiore}(2011)}]{maggiorede}%
  \BibitemOpen
  \bibfield  {author} {\bibinfo {author} {\bibfnamefont {M.}~\bibnamefont
  {Maggiore}},\ }\href {\doibase 10.1103/PhysRevD.83.063514} {\bibfield
  {journal} {\bibinfo  {journal} {Phys. Rev. D}\ }\textbf {\bibinfo {volume}
  {83}},\ \bibinfo {pages} {063514} (\bibinfo {year} {2011})},\ \Eprint
  {http://arxiv.org/abs/1004.1782} {arXiv:1004.1782} \BibitemShut {NoStop}%
\bibitem [{\citenamefont {Paliathanasis}\ \emph {et~al.}(2015)\citenamefont
  {Paliathanasis}, \citenamefont {Pan},\ and\ \citenamefont
  {Pramanik}}]{paliathanasis}%
  \BibitemOpen
  \bibfield  {author} {\bibinfo {author} {\bibfnamefont {A.}~\bibnamefont
  {Paliathanasis}}, \bibinfo {author} {\bibfnamefont {S.}~\bibnamefont {Pan}},
  \ and\ \bibinfo {author} {\bibfnamefont {S.}~\bibnamefont {Pramanik}},\
  }\href {\doibase 10.1088/0264-9381/32/24/245006} {\bibfield  {journal}
  {\bibinfo  {journal} {Class. Quant. Grav.}\ }\textbf {\bibinfo {volume}
  {32}},\ \bibinfo {pages} {245006} (\bibinfo {year} {2015})},\ \Eprint
  {http://arxiv.org/abs/1508.06543} {arXiv:1508.06543} \BibitemShut {NoStop}%
\end{thebibliography}%

\end{document}